\documentclass[useAMS,usegraphicx,usenatbib]{mn2e}

\usepackage{amssymb}
\usepackage{times}

\title[Classical novae from POINT--AGAPE -- II.]{Classical novae from
the POINT-AGAPE microlensing survey of M31 -- II. Rate and statistical
characteristics of the nova population\thanks{Based on observations
made with the Isaac Newton Telescope operated on the island of La
Palma by the Isaac Newton Group in the Spanish Observatorio del Roque
de los Muchachos of the Instituto de Astrofisica de Canarias.}}

\author[M.J.~Darnley et al.]  {M.J.~Darnley$^1$\thanks{E-mail:
mjd@astro.livjm.ac.uk}, M.F.~Bode$^1$, E.~Kerins$^1$, A.M.~Newsam$^1$,
J.~An$^2$, P.~Baillon$^3$,\newauthor 
V.~Belokurov$^2$, S.~Calchi~Novati$^4$, B.J.~Carr$^5$, M.~Cr\'ez\'e$^{6,7}$, N.W.~Evans$^2$, 
\newauthor Y.~Giraud-H\'eraud$^6$,
A.~Gould$^8$, P.~Hewett$^2$, Ph.~Jetzer$^4$, 
J.~Kaplan$^6$, \newauthor S.~Paulin-Henriksson$^6$,
S.J.~Smartt$^{2,9}$, Y.~Tsapras$^5$ and M.~Weston$^5$\\
$^{1}$Astrophysics Research 
Institute, Liverpool John Moores University, Twelve Quays House, Egerton Wharf, 
Birkenhead, CH41 1LD, UK\\
$^2$Institute of Astronomy, University of Cambridge, 
Madingley Road, Cambridge CB3 0HA, UK\\
$^3$CERN, CH-1211 Gen\`eve 23, 
Switzerland\\
$^4$Institut f\"ur Theoretische Physik, Universit\"at Z\"urich,
Winterthurerstrasse 190, CH-8057 Z\"urich, Switzerland\\
$^5$Astronomy Unit, School of Mathematical Sciences, Queen Mary, University of
London, Mile End Road, London E1 4NS, UK\\
$^6$Laboratoire de Physique Corpusculaire et Cosmologie, UMR 7553, CNRS-IN2P3
Coll\`ege de France, 11 Place Marcelin Berthelot, F-75231 Paris, France\\
$^7$Universit\'e Bretagne-Sud, Campus de Tohannic,BP~573, F-56017 Vannes Cedex,
France\\
$^8$Department of Astronomy, Ohio State University, 140 West 18th Avenue,
Columbus, OH 43210, USA\\
$^9$Department of Pure \& Applied Physics, The Queen's University of Belfast, Belfast BT7 1NN, UK
}

\begin{document}
\maketitle

\begin{abstract}
The POINT-AGAPE survey is an optical search for gravitational
microlensing events towards the Andromeda Galaxy (M31).  As well as
microlensing, the survey is sensitive to many different classes of
variable stars and transients.  In our first paper of this series, we
reported the detection of 20 Classical Novae (CNe) observed in Sloan
$r'$ and $i'$ passbands.

An analysis of the maximum magnitude versus rate of decline (MMRD)
relationship in M31 is performed using the resulting POINT-AGAPE CN
catalogue.  Within the limits of the uncertainties of extinction
internal to M31, good fits are produced to the MMRD in two filters.
The MMRD calibration is the first to be performed for Sloan $r'$ and
$i'$ filters.  However, we are unable to verify that novae have the
same absolute magnitude 15 days after peak (the $t_{15}$
relationship), nor any similar relationship for either Sloan filter.

The subsequent analysis of the automated pipeline has provided us with
the most thorough knowledge of the completeness of a CN survey
to-date.  In addition, the large field of view of the survey has
permitted us to probe the outburst rate well into the galactic disk,
unlike previous CCD imaging surveys.  Using this analysis we are able to probe the CN distribution
of M31 and evaluate the global nova rate.  Using models of the
galactic surface brightness of M31, we show that the observed CN
distribution consists of a separate bulge and disk population.  We
also show that the M31 bulge CN eruption rate per unit $r'$ flux is
more than five times greater than that of the disk.

Through a combination of the completeness, M31 surface brightness
model and our M31 CN eruption model, we deduce a global M31 CN rate of
$65^{+16}_{-15}$ year$^{-1}$, a value much higher than found by previous
surveys.  Using the global rate, we derive a M31 bulge rate of
$38^{+15}_{-12}$ year$^{-1}$ and a disk rate of $27^{+19}_{-15}$
year$^{-1}$.  Given our understanding of the completeness and an analysis
of other sources of error, we conclude that the true global nova rate
of M31 is at least $50\%$ higher than was previously thought and this
has consequent implications for the presumed CN rate in the Milky
Way.  We deduce a Galactic bulge rate of $14^{+6}_{-5}$ year$^{-1}$, a
disk rate of $20^{+14}_{-11}$ year$^{-1}$ and a global Galactic rate
of $34^{+15}_{-12}$ year$^{-1}$.
\end{abstract}


\section{Introduction}\label{introdcution}

Classical novae (CNe) undergo unpredictable outbursts with a total
energy that is surpassed only by gamma-ray bursts, supernovae and some
luminous blue variables.  However, CNe are far more commonplace than
these other phenomena \citep{1989clno.book....1W}.

Since their first recorded observations CNe have subsequently been
identified as a sub-class of cataclysmic variables (CVs).  The
canonical model for CVs \citep{1956ApJ...123...44C} is a close
binary system, containing a massive C-O or O-Mg-Ne white dwarf (the
primary) and a low-mass near-main-sequence late-type dwarf that fills
its Roche lobe (the secondary).  Any increase in size through
evolutionary processes of the secondary results in a flow of material
through the inner Lagrangian point into the primary's lobe.  The high
angular momentum of this transferred material causes it to form an
accretion disc around the white dwarf, whilst viscous forces within
the disc act to transfer the accreted material inwards, resulting in a
small amount of the accreted hydrogen-rich material falling on to the
white dwarf's surface \citep{1989clno.book...17K}.  In CN systems the
mass accretion rate is generally lower than $10^{-9}M_{\odot}$
year$^{-1}$ \citep{1998ApJ...496..376C}.  As the accreted layer grows,
the temperature at the base of the material increases.  Hydrogen
burning in the accreted envelope soon develops. Given the correct
conditions, this can lead to a thermonuclear runaway (TNR) in which
the accreted enveloped (and possibly some of the ``dredged-up'' white
dwarf) is expelled from the system in a nova eruption
\citep{1989clno.book...17K,2005clno.book...54S}.

CNe typically exhibit outburst amplitudes of $\sim 10-20$ magnitudes
and display an average absolute blue magnitude of $M_{B}=-8$ at
maximum light, with a limit of around $M_{B}=-9.5$ for the very
fastest \citep{1981ApJ...243..268S,1989clno.book....1W}.  The ability
to accurately measure the distance to many Galactic novae (using
expansion parallax techniques) and a correlation between a nova's
luminosity at maximum light and its rate of decline
\citep{1929ApJ....69..103H,1945PASP...57...69M} makes them potentially
useful as primary distance indicators.  However, until recently,
generally poor light-curve coverage, small sample sizes and a current
lack of understanding of how the properties of CNe vary between
different stellar populations, have severely limited their usefulness
as standard candles.  Nevertheless, their relatively high frequency allows novae to
be used as a tool for mapping the spatial distribution of the
population of close binary systems in nearby galaxies.  CNe may also
be used to test nuclear reaction models and theories, whilst
nucleosynthesis during a nova eruption is thought to make a
substantial contribution to the abundances of a number of chemical
species in the Galaxy such as $^{13}$C, $^{15}$N and $^{17}$O
\citep{2002AIPC..637..104J}.

The ``speed class'' of a CN is often used to describe the overall
timescale of an eruption and to classify a nova
\citep{1939PA.....47..410M,1948AnAp...11....3B}.  The definition of
the various classes depends on the time taken for a nova to diminish
by two (or three) magnitudes below maximum light, $t_{2}$ (or
$t_{3}$).  Throughout this paper we will use the speed class
definitions given in \citet{1989clno.book....1W}.

\subsection {Maximum Magnitude, Rate of Decline relationship}
\label{ss:mmrd}

From his years of observations of CNe in M31,
\citet{1929ApJ....69..103H} noted that the brighter a nova appeared at
maximum the more rapidly its visible light diminished.  Given that
all M31 novae can be considered to lie at equal distance from the
observer, Hubble's observation clearly implied a relationship between
the nova speed class and its maximum magnitude.  These observations
for extragalactic novae were later confirmed for Galactic novae by
\citet{1945PASP...57...69M}, who used a combination of expansion
parallax, interstellar line strengths and Galactic rotation methods to
measure the distances of the nearby novae.  Over time the empirically
determined maximum magnitude versus rate of decline (MMRD) relationship
for CNe has become accepted and refined
\citep{1976A&A....50..113P,1978ApJ...223..351D,1985ApJ...292...90C,2000AJ....120.2007D}.

A recent calibration of the MMRD relationship was made by
\citet{2000AJ....120.2007D} using new distances, derived from
expansion parallaxes, for a sample of 28 Galactic novae, given by

\begin{equation}
M_{V}=(-11.32\pm0.44)+(2.55\pm0.32)\;\mathrm{log}\;(t_{2}\;\mathrm{/days})
\label{empiricMMRD1}
\end{equation}

\noindent\citet{2000AJ....120.2007D} concluded that a linear
relationship is sufficient to model the Galactic MMRD.  They also
derived a typical scatter of $\sim 0.6$ magnitudes for CNe about their
linear fits.  Much of this scatter is thought to be due to
difficulties in measuring accurate distances to the novae
\citep{2000MNRAS.314..175G,2005clno.book....2W,2005clno.book...14S}
and from intrinsic scatter in the optical
decline due to variations in outburst parameters.

However, it is also known that the linear MMRD relationship is not
valid for the fastest and slowest novae
\citep{1956AJ.....61...15A,1957ZA.....41..182S}.  Novae from M31 and
the Large Magellanic Cloud are better described in terms of a
``stretched'' S-shaped curve.  The form is somewhat supported by
theoretical modelling of the nova eruption
\citep{1992ApJ...393..516L}.  The ``flattening'' of the MMRD for
brighter novae is thought to be caused as the mass of the white dwarf
in the central system approaches the Chandrasekhar limit
\citep{1992ApJ...393..516L}.  Conversely, the flattening of the MMRD for the
fainter novae is thought to be an observational selection effect
\citep{1995cvs..book.....W}.

\citet{1989AJ.....97.1622C} were drawn to the conclusion that the same
MMRD relationship is valid in all galaxies of all Hubble types.  This
idea can be exploited to combine data from many different galaxies.
As a result, the MMRD relationship can be used as a fundamental
distance indicator \citep{1981ApJ...243..926S}.  However the use of
the MMRD relationship as a viable distance indicator is dependent upon
being able to accurately measure the maximum brightness of a
particular nova and its speed class, requiring good sampling of both
the maximum light and the decline.

\subsection {Absolute Magnitude 15 days after peak}
\label{ss:t15}

\citet{1955Obs....75..170B} observed that all CNe appeared to reach
approximately the same absolute magnitude  15 days after their maximum
light ($M_{15}$).  The apparent constancy and value of $M_{15}$ is yet
to be fully explained, despite attempts to place it on a more physical
footing \citep{1981ApJ...243..268S}.  Sometimes referred to as the
$t_{15}$ relationship, a recent calibration was carried out by
\citet{2003ApJ...599.1302F} using nine newly discovered novae from
Hubble Space Telescope (HST) observations of M49.  Their calibration
is

\begin{equation}
M_{15,V} =
-6.36\pm0.19\;(\mathrm{random})\pm0.10\;(\mathrm{systematic})
\label{t15cali}
\end{equation}

\noindent However, there is great inconsistency in the calculated
values of $M_{15}$ \citep[see Table~2.4 of][]{2005clno.book....2W}.  More
recent results have called into question the reliability of using this
so-called ``$t_{15}$'' relationship for distance derivations and the validity of the
relationship itself \citep{1992PASP..104..599J,2003ApJ...599.1302F}.

\subsection{CN in M31 and the global nova rate}

A large number of CN surveys in M31 have been carried out, resulting
in the discovery of around 500 novae. These have indicated the global
nova rate in M31 to be $\sim 30-40$~yr$^{-1}$
\citep{2001ApJ...563..749S}.  Table~1 in \citet[][hereafter Paper~1]{2004MNRAS.353..571D}  summarises the
findings of many of these past surveys.  The relatively high nova rate
in M31 and its close proximity to our own Galaxy are major advantages
of targeting M31 for nova surveys.  However, since M31 is nearer
edge-on than face-on, with an inclination angle of $\sim77$\degr\
\citep{1958ApJ...128..465D}, the task of unambiguously distinguishing
between novae erupting within the disk or within the bulge is rather
difficult \citep{1997ApJ...487L..45H}.  Consequently there remains
debate surrounding the distribution and rate of novae within M31.  The
large inclination angle also introduces additional extinction
complications.

The early M31 CNe surveys of \citet{1956AJ.....61...15A} and
\citet{1964AnAp...27..498R} found that the nova distribution decreased
significantly towards the centre of the bulge, with
\citet{1973A&AS....9..347R} reporting the centre of the bulge to be
``{\it devoid of novae}''; all of this was despite their attempts to
detect novae within the central bulge regions.  However, the first M31
H$\alpha$ survey \citep{1987ApJ...318..520C} found that the nova
distribution follows the galactic light all the way into the centre of
the bulge.  A combination of the \citet{1956AJ.....61...15A} novae
with the \citet{1987ApJ...318..520C} catalogue yielded the result that
the bulge nova rate per unit $B$ light was an order of magnitude
greater than that of the disk, implying that the vast majority of the
M31 novae arise from the bulge population.  This result was later
confirmed by \citet{1989AJ.....97.1622C} after undertaking a
comprehensive analysis of all M31 CN data.  However, there is the
potential for biases due to extinction, especially within the disk, as
the H$\alpha$ surveys had focused primarily on the bulge, using much
earlier $B$ band surveys to ``fill in'' the disk data.  In an attempt
to tackle the lingering extinction issues \citet{2001ApJ...563..749S}
extended the H$\alpha$ observations into the M31 disk.  Using M31's
planetary nebula distribution for comparison, they arrived at the
conclusion that the M31 CN distribution is consistent with an
association with the bulge.

\subsection{Nova populations}\label{populations}

The idea that CNe may arise from two distinct populations was first
postulated by \citet{1990LNP...369...34D}.  This was further explored
by \citet{1992A&A...266..232D} who presented evidence that fast novae
were concentrated closer to the Galactic plane than slower novae.
Additional spectroscopic data have revealed that there may exist two
spectroscopic classes of CNe, the Fe {\sc ii} and He/N novae
\citep{1992AJ....104..725W}.  It has been shown the the He/N novae
tend to cluster close to the Galactic plane and that they tend to be
brighter and faster than the Fe {\sc ii} type
\citep{1998ApJ...506..818D}.

Theoretical studies of CN outbursts
\citep[e.g.][]{1980ApJ...239..586S,1981ApJ...243..926S,1982ApJ...257..312P,1992ApJ...393..516L,1995ApJ...445..789P}
have shown that the form of the outburst depends upon properties such
as the white dwarf's mass, accretion rate and luminosity.  These white
dwarf properties may vary with the underlying stellar population.
These findings lend support to the idea that CNe in differing stellar
populations may have distinctly different outburst properties.

The surface gravity of a white dwarf increases with increasing white
dwarf mass.  This leads to a higher pressure at the base of the
accreted envelope when the TNR begins, resulting in a more powerful
outburst.  It also follows that, as the pressure at the envelope base
is greater for more massive white dwarfs, a lower mass of accreted
material is required for the envelope to achieve the temperature and
density required for a TNR to be initiated.  Thus, the more massive
white dwarfs are expected to have shorter recurrence times and to
exhibit faster light-curve evolution.

\subsection{The POINT-AGAPE CNe catalogue}

In Paper~1 we presented an automated pipeline that used objective
selection criteria to detect and classify CNe within a dataset with
good temporal sampling.  We reported 20 CNe erupting within M31 over
three seasons, detected using the pipeline.  Nine of these CNe were
caught during the final rise phase and all were well sampled in at
least two colours.  The excellent light-curve coverage allowed us to
detect and classify CNe over almost the full range of speed classes.

For the purposes of the POINT-AGAPE microlensing survey, the Wide
Field Camera (WFC) on the Isaac Newton Telescope (INT), situated at
La~Palma, was used to regularly monitor two M31 fields between August
1999 and January 2002.  The field centres were located at $\alpha =
0^{\rm h}44^{\rm m}00\fs0$, $\delta = +41\degr34\arcmin00\farcs0$ and
$\alpha = 0^{\rm h}43^{\rm m}23\fs0$, $\delta =
+40\degr58\arcmin15\farcs0$ (J2000)  The WFC consists of a mosaic of
four $2048\times4100$ pixel CCDs and the field locations are indicated
in Figure~1 in Paper~1. The field placements were primarily chosen to
be sensitive to compact dark matter candidates, or Machos, which are
predicted to be most evident towards the far side of the M31 disk
\citep{2001MNRAS.323...13K}.  The observations were conducted over
three seasons in at least two broad-band Sloan-like pass-bands
(usually $r'$ and $i'$, augmented with $g'$ during the first season).
The full distribution of observations can be seen in Table~3 and
Figure~2 in Paper~1.

The outline of the current paper is as follows. In Section~2 we
discuss our treatment of the internal extinction of M31.  Section~3
presents our analysis of the MMRD relationship for M31. The $t_{15}$
analysis is presented in Section~4.  In Section~5 we describe in
detail our completeness analysis of the POINT-AGAPE dataset and the CN
detection pipeline.  In Section~6 we present our analysis of the CN
population of M31.  Section~7 details our evaluation of the global CN
rate of M31 and finally we summarise and discuss our main findings in
Section~8.

\section{Extinction estimation}\label{extinction}

In M31, as with most disk and spiral galaxies, the vast majority of
the dust lies close to the disk plane and the extinction is patchy
\citep{holwerda2004}.  As such, we expect that novae within M31 should
suffer a varying range of extinction, dependent upon their position in
the plane of the galaxy and their line-of-sight displacement.  These
extinction uncertainties may be problematic when trying to analyse
some of the global properties of the POINT-AGAPE CN catalogue, such as
the MMRD and $t_{15}$ relationships, as well as the completeness, CN
population and nova rate.

To compute the extinction across different parts of the M31 disk  we
employ synthetic stellar models (\citeauthor{2001MNRAS.323..109G},
\citeyear{2001MNRAS.323..109G}; Salaris, private communication) to
estimate the extinction free integrated colour ($<r'-i'>$) of the
disk. This is compared to the observed colour across the disk to allow
us to compute an extinction map. Assuming that the M31 reddening curve
is similar to that in the Milky Way, the theoretical and observed
colours yield an extinction map for each band. The global average for
the extinction maps are all re-scaled to give a global average
extinction though the disk of $A(i) = 0.8$, corresponding to the
typical value for Sb galaxies found by \citep[][see their
Figure~11]{holwerda2004} after transforming to Sloan magnitudes
\citep{1998ApJ...500..525S}. Without this rescaling, our extinction
values are systematically high, which we suspect may be because our
synthetic models contain a larger fraction of bluer stars than is
typically present in the M31 disk.

\section{The Maximum Magnitude versus Rate of Decline relationship}\label{mmrd}

As noted earlier, the MMRD relationship is important as it can
potentially be used as a tool to derive the distance to an extragalactic population
of CNe by comparison to the MMRD in our own
Galaxy.

The CN decline rates originally estimated for the 20 detected
POINT-AGAPE novae (see Table~6 of Paper~1) are re-computed for this
analysis.  To calculate the value of $t_{2}$ for
each CN, we linearly interpolate between points on the decline of each lightcurve. This is
carried out for both the $r'$ and $i'$ observations.  The $g'$
observations are omitted as they are only available for seven of the
detected novae.

The measured maximum light and computed $t_{2}$ values for each CN
candidate are shown in Table~\ref{MMRDdat}.  The novae PACN-99-01,
PACN-00-01 and PACN-01-01 are excluded because they are likely
already to have been in decline at first observation; hence it is not
possible to accurately determine the uncertainty in their maximum
light or decline rate.  In addition, only a small portion of the
light-curve of PACN-99-07 is sampled.  As such, due to its erratic
behaviour, we are doubtful whether the classification as a very fast
nova is a true representation of this nova's speed class.  Also, there
are no $i'$ data available around maximum light for PACN-99-05 and
PACN-99-06, so it is not possible to determine a $t_{2}$ value for
these novae.  Therefore, only 16 novae are used for the $r'$ MMRD
analysis and 14 novae for the $i'$ analysis.

\begin{table*}
\begin{minipage}{167mm}
\caption{$r'$ and $i'$ maximum observed magnitudes and corresponding
  $t_{2}$ times, maximum magnitude uncertainties and average
  extinction correction (in magnitudes) for each CN detected in the POINT-AGAPE data.}
\label{MMRDdat}
\begin{tabular}{lllllllll}
\hline
Nova       & $r'(t^{r'}_{0})$   & $t_{2}(r')$  & Estimated max & Estimated    & $i'(t^{i'}_{0})$   & $t_{2}(i')$  & Estimated max & Estimated \\     
           &                    &              & error on $r'$ & average $r'$ &                    &              & error on $i'$ & average $i'$ \\
           &                    &              & maximum light & extinction   &                    &              & maximum light & extinction   \\ 
\hline
PACN-99-01 & $16.53\pm0.03^{a}$ & 30.50        & -             & -0.67        & $16.39\pm0.03^{a}$ & 37.53        & -             & -0.51\\
PACN-99-02 & $18.91\pm0.03$     & 99.48        & -0.03         & -0.57        & $19.19\pm0.04$     & $58.09^{d}$  & -0.04         & -0.43\\
PACN-99-03 & $17.79\pm0.02$     & 59.62        & -0.02         & -0.60        & $17.60\pm0.04$     & 34.16        & -0.06         & -0.46\\
PACN-99-04 & $18.41\pm0.04$     & $164.39^{d}$ & -0.02         & -0.52        & $18.34\pm0.07$     & $87.25^{d}$  & -0.02         & -0.39\\
PACN-99-05 & $17.70\pm0.04$     & 25.82        & -0.04         & -0.66        & -                  & -            & -             & -0.50\\
PACN-99-06 & $16.17\pm0.01$     & 20.30        & -0.13         & -0.67        & -                  & -            & -             & -0.51\\
PACN-99-07 & $18.1\pm0.1$       & 9.80         & -4.2          & -0.65        & $18.02\pm0.04$     & 2.28         & -0.88         & -0.49\\
PACN-00-01 & $17.73\pm0.04^{b}$ & 38.65        & -             & -0.58        & $17.58\pm0.08^{b}$ & $11.88^{d}$  & -             & -0.44\\
PACN-00-02 & $18.15\pm0.03$     & 198.55       & -0.01         & -0.65        & $18.85\pm0.05$     & $817.19^{d}$ & -0.00$^{a}$   & -0.49\\
PACN-00-03 & $18.54\pm0.03$     & 33.02        & -0.06         & -0.67        & $18.19\pm0.04$     & $22.44^{d}$  & -0.09         & -0.51\\
PACN-00-04 & $17.61\pm0.03$     & 30.65        & -0.07         & -0.66        & $17.33\pm0.04$     & $36.44^{d}$  & -0.06         & -0.50\\
PACN-00-05 & $17.30\pm0.01$     & 59.21        & -0.18         & -0.65        & $17.11\pm0.01$     & 198.32       & -0.06         & -0.49\\
PACN-00-06 & $17.09\pm0.01$     & 13.85        & -0.09         & -0.65        & $16.64\pm0.01$     & 13.44        & -0.14         & -0.49\\
PACN-00-07 & $19.53\pm0.04$     & 55.21        & -0.03         & -0.48        & $19.48\pm0.05$     & $100.27^{d}$ & -0.02         & -0.37\\
PACN-01-01 & $18.45\pm0.02^{c}$ & $213.12^{d}$ & -             & -0.61        & $18.16\pm0.04^{c}$ & $330.86^{d}$ & -             & -0.47\\
PACN-01-02 & $17.14\pm0.03$     & 22.06        & -0.10         & -0.64        & $16.71\pm0.04$     & 17.08        & -0.12         & -0.48\\
PACN-01-03 & $17.30\pm0.04$     & 143.71       & -0.02         & -0.62        & $16.88\pm0.06$     & $66.21^{d}$  & -0.03         & -0.47\\
PACN-01-04 & $17.90\pm0.03$     & 47.29        & -0.05         & -0.65        & $17.38\pm0.04$     & 37.24        & -0.06         & -0.49\\
PACN-01-05 & $15.90\pm0.01$     & 28.15        & -0.93         & -0.57        & $15.61\pm0.01$     & 15.83        & -1.68         & -0.43\\
PACN-01-06 & $17.38\pm0.01$     & 52.13        & -0.12         & -0.63        & $16.88\pm0.03$     & $38.24^{d}$  & -0.10         & -0.48\\
\hline
\end{tabular}

\medskip
$^{a}$ the light-curve of PACN-99-01 was visible at, or shortly after,
maximum light in the first observational epoch of the first season,
so it was only possible to put a lower limit on its maximum light.
$^{b}$ as PACN-99-01 for the second season.  $^{c}$ as PACN-99-01 for
the third season.  $^{d}$ it was not possible to follow these
light-curves through two magnitudes below their observed maxima, so
the value of $t_{2}$ has been estimated from the general trend of
these light-curves.

\end{minipage}
\end{table*}

An initial evaluation of the linear MMRD relationship for both the
$r'$ and $i'$ data produces a poor fit, in the sense that the scatter of both
distributions is much greater than that implied solely by the
photometric uncertainties.

\subsection{Maximum light uncertainties}

In an attempt to further refine our MMRD relationship, and either
reduce or help to explain the large scatter, we allow that
the brightest observation of each nova is only a lower limit to that
nova's true maximum light.  In the majority of cases each maximum
observation is straddled by observations from the following and
preceding nights, leading to a small error in the assignment of the
true maximum light.  We then estimate the maximum potential error on
our measurement of the maximum light.  Taking a good estimation of the
general slope of each light-curve to be $2/t_{2}$ magnitudes/day, we
calculate the amount that each light-curve could have possibly
increased in brightness between the two points straddling the
brightest observation.  The maximum potential error induced by missing
the maximum light for each CN is also shown in Table~\ref{MMRDdat}.  The
large maximum light errors derived for PACN-99-07 and PACN-01-05 are
due to a combination of a fast decline rate and poor sampling around
maximum light.  In particular, the speed class assignment of
PACN-99-07 is known to be suspect; a much slower decline time
seems more fitting.  As previously mentioned, this CN is excluded
from this analysis due to the uncertainty of its decline rate.

The photometric and maximum light uncertainties are combined by simple
addition as the maximum light error is a systematic rather than a
random error.  The actual maximum light is equally likely to lie at
any point between the observed value and the computed potential
maximum value.  For the purpose of fitting the MMRD relationship, we
``re-sample'' the ``observation'' of the maximum light to be the
mid-point of the computed range for each CN.

We use a minimum absolute deviation method to fit the data, as our
errors are no longer Gaussian and are dominated by uniformly
distributed systematic uncertainties.  The scatter on these MMRD fits
remains at $\sim0.6$ and $\sim0.7$ magnitudes for the $r'$
and $i'$ data respectively.  This implies that the scatter in the MMRD
relationships is not significantly affected by missing the maximum by
a day or two, as is the typical interval expected in the POINT-AGAPE
data.  The similarity between these fits and those performed without
considering the maximum light uncertainties is also indicative of the
apparent minimal effect induced by considering the possibility of
missing the maximum light of a CN.

\subsection{Extinction corrections}
Each CN's light-curve may still be affected by extinction within M31
and our own Galaxy.  The Galactic extinction in the direction of M31
is well defined and relatively small compared to the maximum potential
extinction experienced within M31.  The extinction experienced by each
nova's light depends upon the column density between the nova and the
observer, with the maximum potential extinction dependent upon the
CN's position.

\begin{figure*}
\rotatebox{270}{\includegraphics[width=230pt]{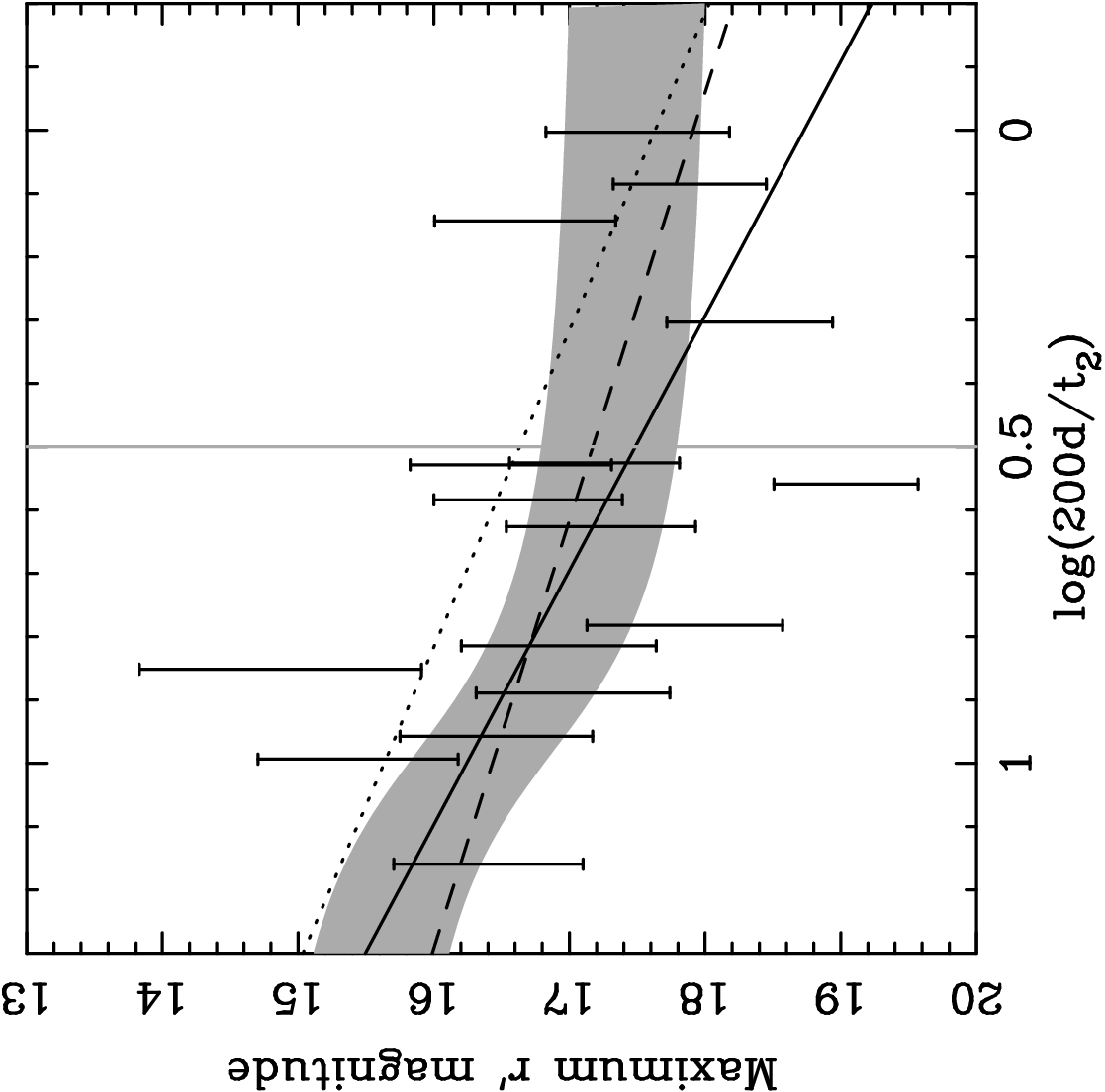}}
\rotatebox{270}{\includegraphics[width=230pt]{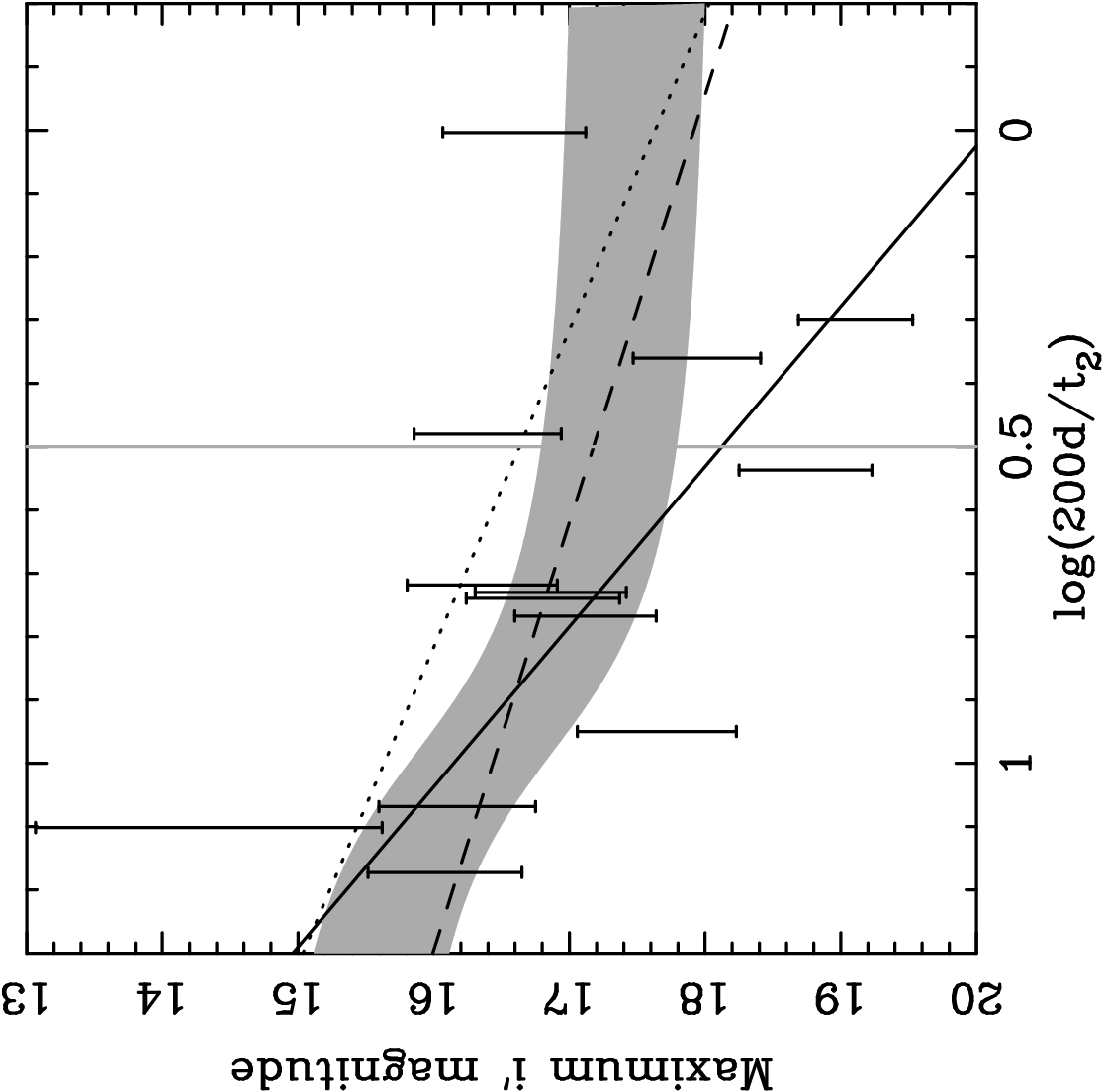}}
\caption{The relationship between the $r'$ brightness at
  maximum light and the decay rate ($v_{2}(r')=\log[200d/t_{2}(r')]$)
  of the 16 POINT-AGAPE CNe with well defined maximum lights and decay
  rates (left pane).  Likewise for $i'$ maximum light (right pane).  The range of the error bars represents the maximum estimated
  range of the combination of the extinction, maximum light and
  observational uncertainties.  The dashed line indicates an unweighted
  fit performed on all the data (Equation~\ref{calcMMRD3}), while the
  solid line shows an unweighted fit performed on the data in the
  range $0.5\leq\log[200d/t_{2}(r')]\leq1.5$ (see
  Equation~\ref{calcMMRD3b}).  The vertical grey line represents the
  ``slow'' boundary of the linear region of the MMRD
  ($\log[200d/t_{2}(r')]=0.5$).  The grey shaded region represents the
  best fit Galactic ``S-shaped'' MMRD, and the black dotted line shows the
  best fit Galactic linear MMRD - both these Galactic MMRD are derived
  for $V$ data and have been transformed to the M31 distance.}
\label{MMRD-plot}
\end{figure*}

We use the extinction maps (see Section~\ref{extinction}) to provide
an estimate of the maximum potential extinction experienced by each of
the 16 CNe for which valid $r'$ decline rates could be computed and
by each of the 14 novae with valid $i'$ decline rates.  The computed average
extinctions are shown in Table~\ref{MMRDdat}.  Figure~\ref{MMRD-plot}
shows the distribution of extinction-corrected and maximum-magnitude
corrected $r'$ and $i'$ data respectively.  The base of the
error bars represents the true maximum observed light, with the length
of the bars representing the absolute range of the actual maximum
light.  As with the maximum magnitude errors, we assume that the
extinction is equally likely to lie at any value between zero and the
maximum estimate.  Again, for this analysis, the three error sources
are combined by simple addition.  As the maximum light error and the
extinction error are both independent absolute maximum errors, the
maximum error that can be experienced due to both of these sources is
simply the sum of the two -- in the direction of increasing
luminosity.  We again assume that the best-guess maximum-light flux occurs
equally distant between the observation and the extreme maximum error
for the purpose of the MMRD fitting.  The fits to the maximum-light estimates are
\begin{equation}
m_{r'}=(14.5\pm1.3)+(1.5\pm0.8)\;\mathrm{log}\;(t_{2}\;\mathrm{/days})
\label{calcMMRD3}
\end{equation}
\begin{equation}
m_{i'}=(14.5\pm1.0)+(1.5\pm0.6)\;\mathrm{log}\;(t_{2}\;\mathrm{/days})
\label{calcMMRD3_i}
\end{equation}
 The scatter in the final MMRD fits is $\sim0.7$ and
$\sim0.8$ magnitudes for the $r'$ and $i'$ datasets respectively,
comparable to the mean error size.  The MMRD data are also analysed
in the linear region of the ``S-shaped'' curve
($0.5\leq\log[200d/t_{2}]\leq1.5$), yielding
\begin{equation}
m_{r'}=(13.0\pm2.2)+(2.5\pm1.4)\;\mathrm{log}\;(t_{2}\;\mathrm{/days})
\label{calcMMRD3b}
\end{equation}
\begin{equation}
m_{i'}=(11.0\pm2.4)+(3.9\pm1.6)\;\mathrm{log}\;(t_{2}\;\mathrm{/days})
\label{calcMMRD3b_i}
\end{equation}
with a scatter of $\sim0.7$ magnitudes about the fits for
both the $r'$ and $i'$ data, which is again comparable to the mean
error.

All four of the MMRD relationships calculated
(Equations~\ref{calcMMRD3}-\ref{calcMMRD3b_i}) are shown in
Figure~\ref{MMRD-plot}, along with a recent Galactic calibration of the
MMRD relationship (Equation~1 in
\citealt{2000AJ....120.2007D}).  The \citet{2000AJ....120.2007D}
calibration has been translated to the distance of M31 using a
distance modulus of 24.3 magnitudes \citep{1986ApJ...305..583W},
yielding

\begin{equation}
m_{V}=(12.98\pm0.44)+(2.55\pm0.32)\;\mathrm{log}\;(t_{2}\;\mathrm{/days}).
\label{empiricMMRD-M31}
\end{equation}

It is clear that neither the $r'$ or $i'$ MMRD relationships provide
enough data points to investigate in greater detail the true form (linear of S-shaped)
of the M31 MMRD relationship.  As previous work (see
Section~\ref{ss:mmrd}) has specifically identified a linear region
within M31's MMRD relationship, we will use the relationships defined
within the linear region (shown in Equations~\ref{calcMMRD3b} and
\ref{calcMMRD3b_i}) as the $r'$ and $i'$ MMRDs for M31.

We note that extinction is a rather weak factor in determining the slope of the MMRD in the linear regime, where all the novae are intrinsically very bright $[M(r') lta 17]$. Indeed our MMRD slope determination with or without the extinction correction is essentially the same.

\section{The \lowercase{\boldmath{$t_{15}$}} relationship}\label{15}

As discussed earlier, the $t_{15}$ relationship may also be useful
in calculating the distance to a CN population.  The majority of
previous $t_{15}$ calibrations have been carried out using $V$-band
data.  However, due to the restrictions of the POINT-AGAPE catalogue,
we can only attempt calibration using $r'$ and $i'$ data.

Using these data, an initial calibration of the $t_{15}$ relationship
for the POINT-AGAPE catalogue indicates that the photometric errors
alone do not account for the extent of the scatter of luminosities at 15
days following maximum light.

As was attempted for the MMRD data, we can try to decrease or at
least explain the scatter in these data by taking into account the
line-of-sight extinction.  Using the data in
Table~\ref{MMRDdat} to recalibrate each light-curve, we
therefore reassess both the $r'$ and $i'$ $t_{15}$ relationships
for the POINT-AGAPE novae catalogue:

\begin{equation}
m_{15,r'}=18.0\pm0.9
\label{t15_eq_r_2}
\end{equation}

\begin{equation}
m_{15,i'}=18.0\pm1.0.
\label{t15_eq_i_2}
\end{equation}

\begin{figure*}
\rotatebox{270}{\includegraphics[width=230pt]{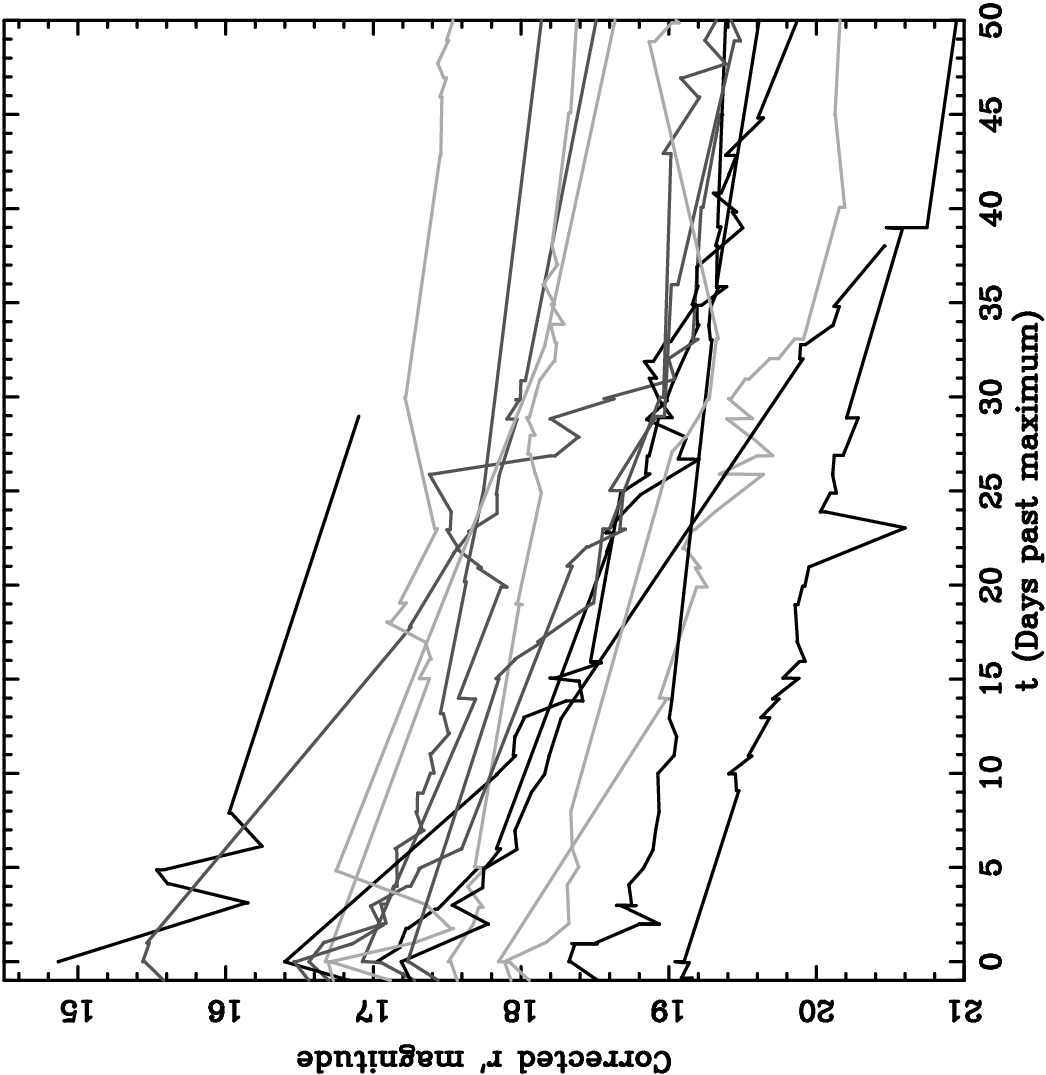}}
\rotatebox{270}{\includegraphics[width=230pt]{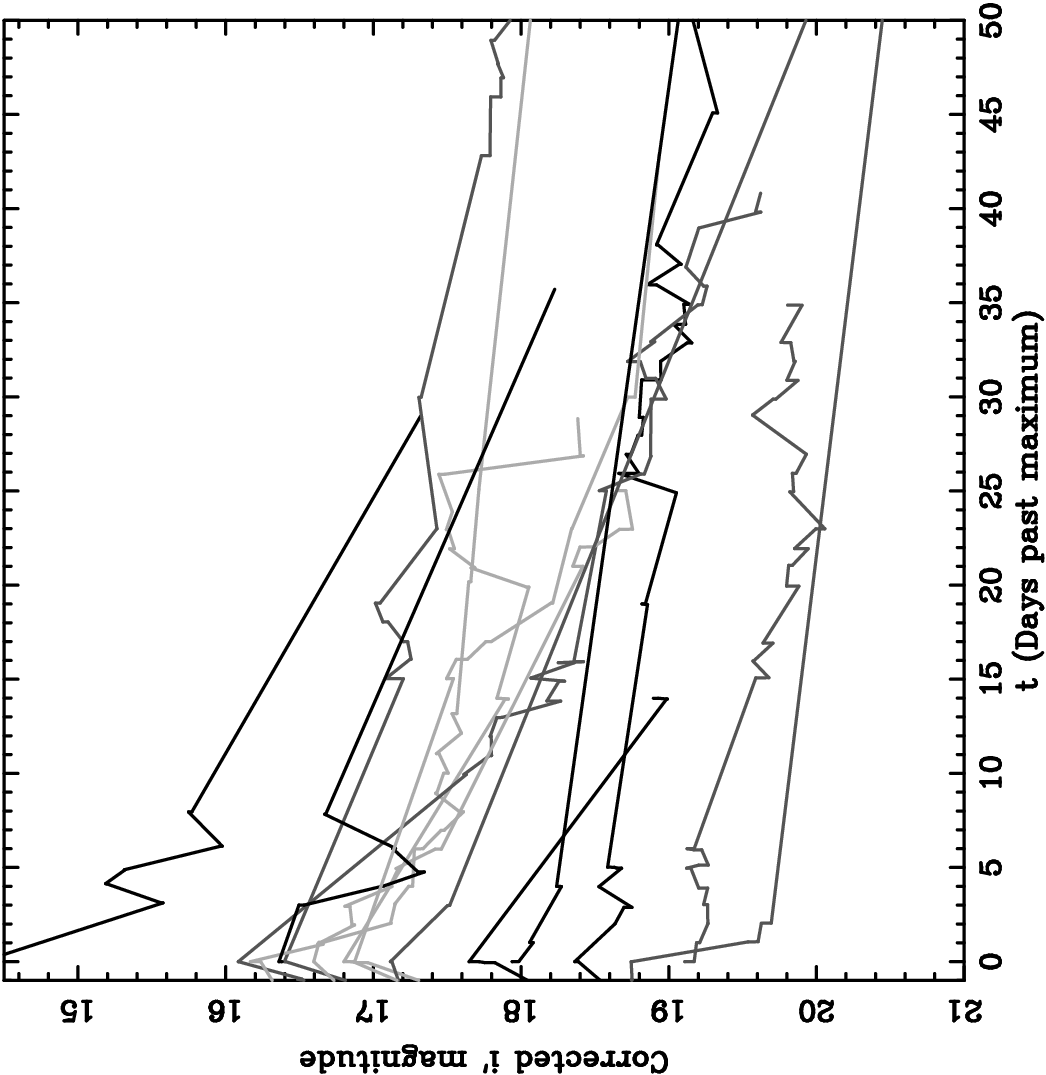}}
\caption{A superposition of the recalibrated $r'$ light-curves of 16
  of the POINT-AGAPE CNe (left pane) and the superposition of the
  recalibrated $i'$ light-curves of 14 of the POINT-AGAPE CNe (right pane).  The light-curves have been time-shifted so
  that the times of their observed maximum light are coincident.  Each
  line represents the linear interpolation of the light-curves between
  observations.}
\label{t15}
\end{figure*}

\noindent Figure~\ref{t15} shows superpositions of the 16 $r'$ and 14
$i'$ re-calibrated light-curves that have well defined maximum light
magnitudes.  The light-curves are all plotted in units of time since
maximum light.  Each plot shows the light-curve behaviour for the
first 50 days following each eruption.  There is clearly little or no
convergence of these light-curves at time around 15 days.

The inclusion of extinction and maximum light uncertainties has
actually slightly increased the scatter of the $t_{15}$ values.  This
implies that the scatter in luminosities 15 days after peak is not
solely due to uncertainties in the luminosity of the nova,
indicating that any $t_{15}$ relationship may not be valid for these
pass bands.

\subsection{Comparison with previous results}

By assuming a distance modulus for M31 of 24.3 magnitudes
\citep{1986ApJ...305..583W}, our computed $t_{15}$ values (see
Equations~\ref{t15_eq_r_2} and \ref{t15_eq_i_2}) might be compared
with those found previously \citep[see Table~2.4 of][]{2005clno.book....2W}:
\begin{equation}
M_{15,r'}=-6.3\pm0.9
\label{t15_eq_r_3}
\end{equation}
\begin{equation}
M_{15,i'}=-6.3\pm1.0.
\label{t15_eq_i_3}
\end{equation}
However, direct comparison between our results and previous results
can not be easily made.  All calibrations of the $t_{15}$ relationship
to-date have been in ``blue'' bands, whereas our calibration is
performed in ``red'' bands.  Nevertheless, given that CNe are expected to
become bluer with time, the result that our $t_{15}$ luminosities are
fainter than all but the most recent of the previous ``blue''
calibrations \citep[see Table~2.4 of][]{2005clno.book....2W} is not
surprising.  The scatter in our results is very large, and larger than
those found in previous surveys.  For instance, in a recent HST study
of M49 \citep{2003ApJ...599.1302F} the $V$-band $t_{15}$ relationship
was found to have $\sigma=0.43$ magnitudes. It should also be noted
that, as an elliptical galaxy, M49 does not suffer from large
extinction problems.  However, the large degree of the scatter in the
POINT-AGAPE data can not be solely explained by the combination of
maximum light and extinction uncertainties, whose mean error is
$\sim0.7$ magnitudes for both the $r'$ and $i'$ data.

\begin{figure}
\rotatebox{270}{\includegraphics[width=230pt]{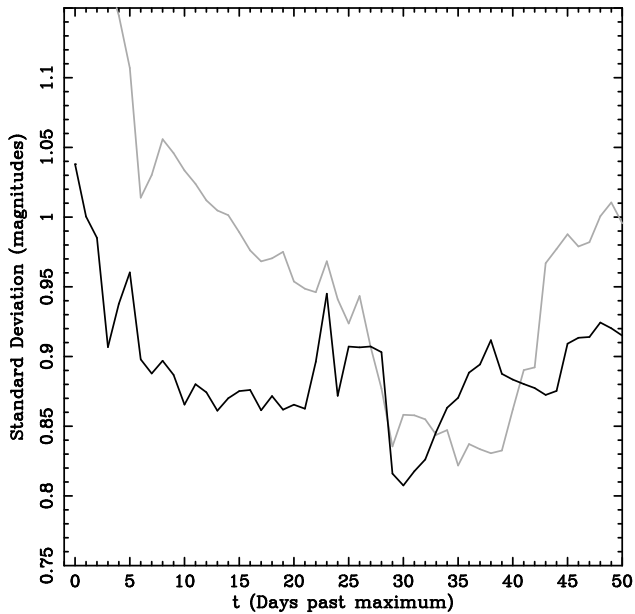}}
\caption{Plot of the distribution of $r'$ magnitude scatter (black
  line) and the $i'$ magnitude scatter (grey line) between observed
  nova magnitudes for a range of times following maximum light.}
\label{t15_scatter}
\end{figure}

As a final test, if the $t_{15}$ relation is valid, then we would
expect a minimum in the scatter of the light-curves at or around 15
days after maximum light.  Figure~\ref{t15_scatter} shows a plot of
the scatter between the POINT-AGAPE light-curves over a large range of
time following maximum light for the $r'$ and $i'$ data.  From
inspection of this plot, it is quite clear that the light-curves of the
POINT-AGAPE sample do not exhibit behaviour consistent with the
existence of a $t_{15}$ relationship.  However, the $r'$ scatter does
seem to exhibit a minimum at $\sim30$ days after maximum light and the
$i'$ scatter is minimised $\sim35$ days following maximum.  However,
these minima are coincident with the end of data sampling for a number
of the light-curves, so may just be indicative of the temporal
coverage of the POINT-AGAPE light-curves themselves.

\section{CN detection pipeline completeness}\label{completeness}

Because we have have selected our nova candidates using objective selection criteria, we
can assess the efficiency of our selection pipeline. This allows us to compute 
the completeness of the POINT-AGAPE CN catalogue and aids us in
probing the underlying CN distribution and to compute a robust estimate of the
global nova rate.  To measure the completeness of the catalogue, we seed
the raw POINT-AGAPE data with re-sampled light-curves of our 20
detected CNe.  We then re-run the entire CN detection pipeline on
these seeded data to allow us to compute the proportion of recovered
light-curves.

\subsection{Creating test light-curves}

The seeded light-curves are positioned on a grid within the aligned
image data stack (see Section~3 in Paper~1), with a grid spacing of 15
pixels (5 arcmin).  This grid spacing is chosen to allow the closest
possible spacing of seeded objects, whilst minimising overlap of each
star's point spread function (PSF).  Each light-curve is seeded at a
random eruption epoch, such that at least one point of the light-curve
occurred between the first observational epoch and the final epoch.

In order to seed the detected novae at any random time we linearly
interpolate their light-curves between successive observations.
Whilst this works well when the timescale between observations is
small, it becomes less reliable when the gaps are larger.  The largest
gaps in the observations are usually of order 2 weeks, but in a few
cases, when light-curves are followed across two seasons, these are up
to six months.  The two light-curves affected by this are PACN-00-02
and PACN-00-05 (see Section~5 of Paper~1), these light-curves being
linearly interpolated across the seasonal gaps.  Given the form of the
light-curve of PACN-00-02, we expect this method to be relatively
reliable as an estimate of the flux.  However, given the predicted
transition phase minimum for PACN-00-05, this estimate is much less
reliable.

\subsection{Seeding the raw POINT-AGAPE data}

The generated light-curves are added to both the raw data and the
PSF-matched data (see Section~3.2 of Paper~1) using the NOAO IRAF
package environment\footnote{IRAF is distributed by the National
Optical Astronomy Observatories, which are operated by the Association
of Universities for Research in Astronomy, Inc.  under cooperative
agreement with the National Science Foundation.} {\tt mkobjects},
which scales the relevant image's PSF profile to the correct
luminosity, then adds the scaled PSF profile to the data, recalculates
the Poisson errors and combines these with the data.

\subsection{Re-running the nova detection pipeline}

The CN detection pipeline is re-run on the seeded data; however a
number of stages of the pipeline are not used.  As both the raw and
the PSF-matched data are seeded independently\footnote {In order to
maintain consistency between the seedings in the raw and PSF-matched
data, the same random seed is used to regenerate the Poisson noise
for both CN seedings.}, the image alignment, trimming, PSF-matching
and background estimate stages are not required.

The seeded PSF-matched data are run though the aperture photometry
pipeline (see Sections~4.2 and 4.3 of Paper 1) to produce the
preliminary list of recovered light-curves.  PSF-fitting photometry is
then performed at the position of each of the seeded novae recovered
from the aperture photometry stage.  These nova light-curves are then
passed back through the ``peak detection'' stage of the pipeline.
However, all of the pipeline stages that are related to the colour
light-curves are ignored.  We are able to ignore the colour criteria
as these are solely introduced to distinguish CN light-curves from
the light-curves of other objects that may have passed through the
previous stages of the pipeline\footnote {The make-up of the
POINT-AGAPE observation strategy makes it impossible to seed $i'$ and
$g'$ band data across observing seasons as there is minimal $i'$ band
data available for the 1st season and no $g'$ band data available for
the 2nd or 3rd.}.  However, as we know that all 20 of the seeded
light-curves are those of CN discovered in the POINT-AGAPE data, they
have already passed the colour criteria.

A seeded CN light-curve may ``fail'' the pipeline for any of the
following reasons:

\begin{enumerate}
{\item The object has been seeded at a location in the M31 field
  where, due to the brightness of the background and/or surrounding
  objects, it is impossible to make a $10\sigma$ detection of the
  object at any epoch.}   
{\item There are not five consecutive
  detections, either because the object is too faint to detect or
  because it has been seeded such that there are not five observations
  in which the nova was visible.}   
{\item The observed ``peak'' of
  the seeded nova is not significant enough to pass the primary peak
  test.  This is either due to the galactic background or because the
  nova has been seeded ``low down'' in its light-curve, i.e. the
  actual peak has not been seeded in any of the observations.}  
{\item
  A seeded CN light-curve can fail the periodicity test, the secondary
  peak height test or the ``$<90\%$ of points in peaks'' test, if the
  nova has been seeded close to a region of the image that also varies
  significantly with time.  This may be due to a nearby variable star,
  a region of bad pixels or a saturated object.  Again, it is possible
  for a light-curve to fail this test if data around the actual peak
  of the nova has not been seeded.}
\end{enumerate}

The numerical results of the completeness run of the CN pipeline are
shown in Table~\ref{stat}.

\begin{table*}
\begin{minipage}{144mm}
\caption{The effect of each stage of our selection pipeline upon the
synthetic classical nova catalogue.  These steps are described in Section~4 of Paper~1.}
\label{stat}
\begin{tabular}{llllllllll}
\hline
Pipeline                   & \multicolumn{4}{c}{North-field}      & \multicolumn{4}{c}{South-field}      & All\\
stage                      & CCD1    & CCD2    & CCD3    & CCD4   & CCD1    & CCD2    & CCD3    & CCD4   & CCDs\\
\hline
Seeded objects             & 35,239  & 35,376  & 35,244  & 35,376 & 35,910  & 35,239  & 35,378  & 34,864 & 282,626\\
Objects seeded within data & 18,291  & 18,886  & 18,314  & 18,401 & 14,185  & 18,082  & 18,861  & 18,527 & 143,547\\
$10\sigma$ objects         & 16,519  & 17,569  & 17,339  & 17,107 & 13,352  & 16,216  & 17,667  & 17,049 & 132,818\\
\multicolumn{9}{c}{{\em Pipeline $1st$ pass -- aperture photometry}}\\
5 consecutive detections   & 13,457  & 14,982  & 14,945  & 14,428 & 12,451  & 14,573  & 15,849  & 15,303 & 115,988\\
$\geq1$ primary peak       & 13,078  & 14,768  & 14,660  & 14,106 & 12,221  & 14,196  & 15,409  & 14,789 & 113,227\\
Periodicity test           & 12,979  & 14,676  & 14,550  & 13,965 & 12,156  & 14,118  & 15,191  & 14,583 & 112,218\\
Primary peak height        & 12,859  & 14,564  & 14,477  & 13,784 & 12,113  & 13,945  & 14,938  & 14,391 & 111,071\\
Secondary peak height      & 11,011  & 11,642  & 12,460  & 11,253 & 10,373  & 11,926  & 11,062  & 11,179 & 90,906\\                     
\multicolumn{9}{c}{{\em Pipeline $2nd$ pass -- PSF-fitting photometry}}\\
5 consecutive detections   & 9,981   & 11,549  & 12,330  & 11,191 & 10,157  & 11,502  & 10,628  & 10,787 & 88,125\\
$\geq1$ primary peak       & 9,083   & 11,056  & 11,907  & 10,696 & 9,728   & 10,837  & 9,978   & 10,157 & 83,442\\
Periodicity test           & 9,083   & 11,056  & 11,907  & 10,696 & 9,728   & 10,837  & 9,978   & 10,157 & 83,442\\
Primary peak height        & 8,868   & 10,805  & 11,531  & 10,420 & 9,461   & 10,591  & 9,805   & 9,909  & 81,390\\
Secondary peak height      & 8,169   & 10,433  & 10,994  & 9,938  & 9,290   & 10,209  & 9,619   & 9,682  & 78,334\\
\multicolumn{9}{c}{{\em Further candidate elimination stages}}\\
$<90\%$ of points in peaks & 8,141   & 10,403  & 10,968  & 9,911  & 9,274   & 10,185  & 9,605   & 9,654  & 78,141\\
5 $g'$ or $i'$ points      & -       & -       & -       & -      & -       & -       & -       & -      & -\\
Colour evolution           & -       & -       & -       & -      & -       & -       & -       & -      & - \\
Rate of decline            & -       & -       & -       & -      & -       & -       & -       & -      & - \\
Colour--magnitude criteria & -       & -       & -       & -      & -       & -       & -       & -      & - \\
\\
{\bf Final candidates}     & {\bf 8,141} & {\bf 10,403} & {\bf 10,968} & {\bf 9,938} & {\bf 9,274} & {\bf 10,185} & {\bf 9,605} & {\bf 9,654} & {\bf 78,141}\\
\hline
\end{tabular}
\end{minipage}
\end{table*}

\subsection {Completeness distribution}
\label{se:comp}

In order to compute the completeness, we subdivide each CCD into 1
arcmin grid squares, with each grid square containing 144 novae seed
points.  Depending upon the size of the trimmed CCDs (see Section~3.1
of Paper~1), the CCDs contain between 200 and 242 grid squares.

We first compute the completeness distribution of the CN pipeline;
this distribution tells us the probability of the pipeline detecting a
CN -- of a type originally detected by the pipeline -- at any position
in the POINT-AGAPE fields, given that the CN in question is
``visible'' at least once between (and including) the first and last
observations.  This completeness calculation takes account of all of
the major factors that affect the completeness of the detection
pipeline, including the temporal distribution of the POINT-AGAPE
observations, the galactic surface brightness, the variety of CN
light-curve forms and any interference by foreground objects.

\begin{figure}
\rotatebox{270}{\includegraphics[height=230pt]{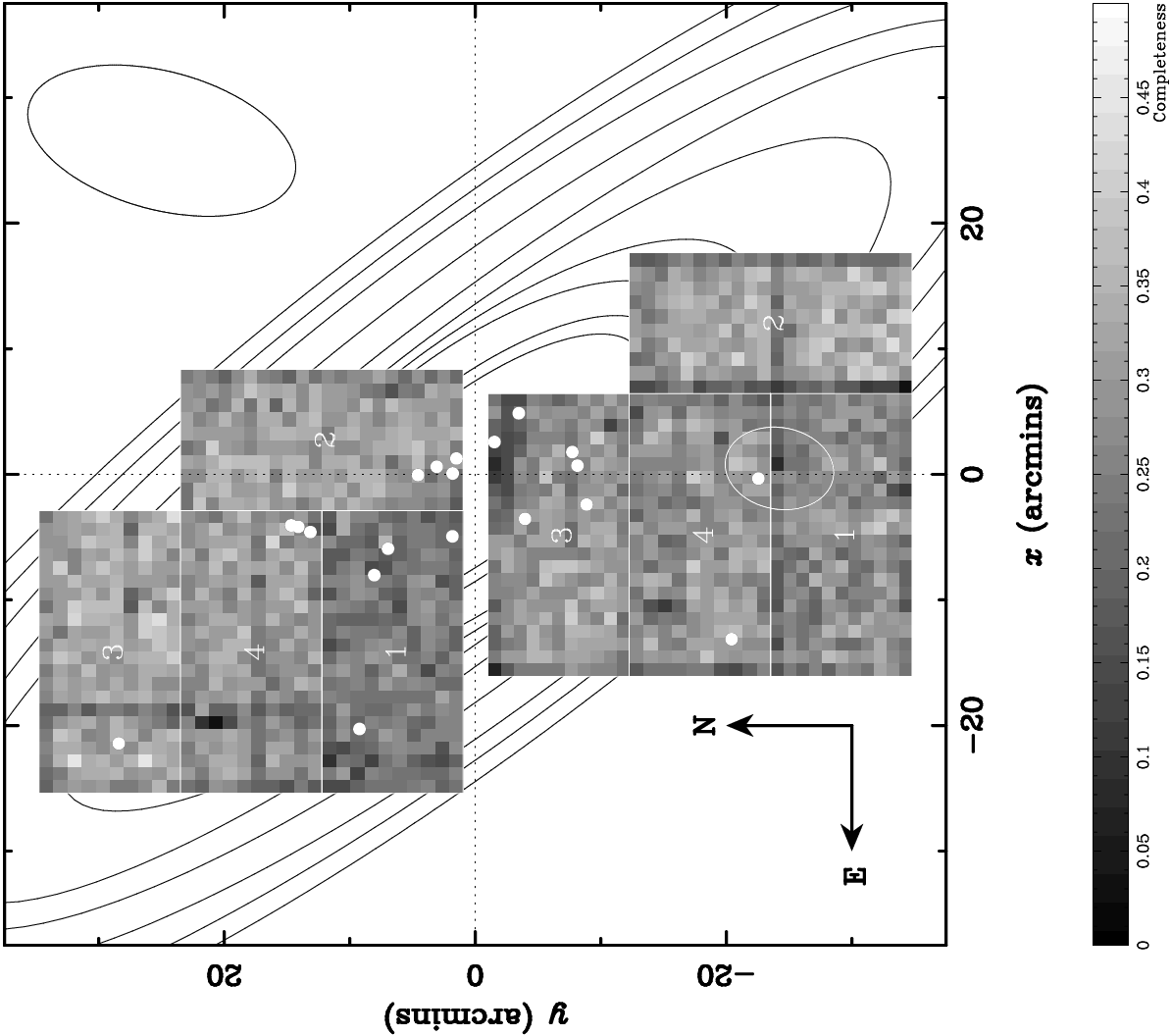}}
\caption{The POINT-AGAPE CNe detection pipeline completeness
  distribution.  Each rectangle represents one of the four INT WFC
  CCDs and the white circles indicate the positions of the 20 detected
  novae.  The origin is the centre of M31 at (J2000) $\alpha = 0^{\rm
  h}42^{\rm m}44\fs324$, $\delta = +41\degr16\arcmin08\farcs53$
  \protect\citep{1992ApJ...390L...9C}. Also indicated are ten
  representative M31 ``isophotes'' from the surface photometry of
  \protect\citet{1958ApJ...128..465D}, along with representations of
  the positions and sizes of M32 (within the southern field) and
  NGC205.  }
\label{complete}
\end{figure}

The generated completeness map is shown in Figure~\ref{complete}.
This illustrates that the completeness is relatively flat across both
fields, within the noise, at a value between about
$30\%-40\%$. However the completeness does decline towards the centre
of the galaxy, as the galactic background begins to increase
significantly.

\section{M31 CN population}\label{population}

If we adopt the simplest assumption, that the nova distribution in M31
follows the light distribution \citep{1987ApJ...318..520C} then the
probability of a CN erupting at a particular point within M31 is
proportional to the flux at that position.  By requiring that a given
CN erupts within one of the two POINT-AGAPE fields, we can compute the
probability of a CN erupting at a particular point within M31.
However, from our completeness calculations, we also know the
probability of detecting a CN in each grid square, given that a nova
erupts within that square.  Hence we can use this to calculate the
probability of detecting a CN within the POINT-AGAPE fields, given
that a nova erupts within them.  This detection probability is given
by:

\begin{equation}
P_{i}=\frac{f_{i}}{\sum_{j=1}^{N_{bins}}f_{j}}\cdot\varepsilon_{i}
\label{probdist}
\end{equation}

\noindent where $P_{i}$ is the probability of a CN erupting in a
particular grid square containing a flux $f_{i}$ and $\varepsilon_{i}$
is the computed pipeline efficiency in the grid square in question.
Because of the uniformity of $\varepsilon$, the distribution of
detection probabilities closely resembles the galactic light.
However, as the completeness drops slightly towards the galactic
centre, this distribution has a slightly weaker central dependence
than the flux.

If M31 consisted solely of a single population of stars, then the
``nova follows the light'' distribution would be a good model of the
CN distribution.  However, with recent evidence pointing towards
separate bulge and disk populations of novae (see
Section~\ref{populations}), it is likely that the detection probability
model (Equation~\ref{probdist}) needs to be modified.

\subsection {Modelling M31's galactic light}\label{model}

In order to calculate the overall CN rate of M31 or to investigate the
possibility of separate bulge and disk CN populations, we need to be
able to compute both the bulge and disk component of the light at any
given point within the galaxy.  In order to do this, we need to create
a model of the flux distribution of M31.  To perform this
modelling we subdivide each CCD using the one arcmin square grid
system that we employed for the completeness calculations.

In order to try to fit the disk or bulge components of the M31
light, we first define a region of the galaxy within which either the
bulge or the disk light could be unambiguously defined.  In the outer
regions of a spiral galaxy such as M31, the visible light arises
almost completely from the disk.  So it is possible to model the disk
in these regions and extend the model to the inner regions of the
galaxy.

To greatly simplify the geometry of the galactic disk, we make the assumption that it is 
thin with an inclination of 77\degr\ \citep{1958ApJ...128..465D}
and that the flux distribution is smooth across the disk.  Another
simplification we make is to essentially collapse the disk into a
one-dimensional system.  Each position within the disk is transformed
to the semi-major axis of the ellipse that passed through that point.
As the light from a galactic disk can often be modelled using a simple
exponential law \citep{1970ApJ...160..811F}, the disk flux at any
point within M31 can then be defined as
\begin{equation}
f_{d}(a_{d})=f_{d}^{0}\;\mathrm{e}^{-a_{d}/a_{d}^{0}}
\label{diskmodel}
\end{equation}
where $f_{d}(a_{d})$ is the disk flux at a position within
the disk with semi-major axis $a_{d}$.  As it is not possible to
unambiguously separate disk light from bulge light, the fit is
performed to the total flux data for $a_{d}\geq40$ arcmin in order to
have minimal contamination from the bulge light.  The best-fit values
found for the two parameters are, $f_{d}^{0}=3,676$ adu/pixel and
$a_{d}^{0}=43.1$ arcmin.

In order to attempt to fit the bulge flux, we extend the disk model
across the whole galaxy and then subtract the modelled disk light
from the galactic light to leave just the bulge light and the
residuals to the disk model.  We model the bulge with elliptical
isophotes with an axis ratio $b/a=0.6$ \citep{1987ApJ...318..520C}. We
again transform the spatial positions of each point within the bulge
to the semi-major axis of the ellipse on which that point lies.  We
produced the following bulge model using a standard $r^{1/4}$ law
\citep{1948AnAp...11..247D,1953MNRAS.113..134D}:
\begin{equation}
\mathrm{log}[f_{b}(a_{b})/f_{b}^{0}]=-3.33[(a_{b}/a_{b}^{0})^{1/4}-1]
\label{bulgemodel}
\end{equation}
where $f_{b}(a_{d})$ is the bulge flux at a position within
the bulge with semi-major axis $a_{b}$.  A fit is performed to the
bulge flux data for $a_{b}\leq15$ arcmin, so that the fit is not
influenced by the disk fit residuals.  The best-fit values are
$f_{b}^{0}=6,914$ adu/pixel and $a_{b}^{0}=5.1$ arcmin.

\begin{figure}
\rotatebox{270}{\includegraphics[width=230pt]{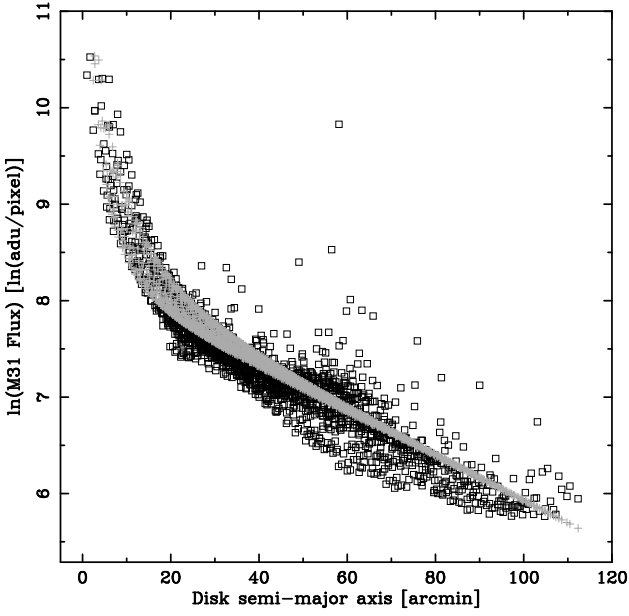}}
\caption{A plot of
  M31 flux against disk-semi-major axis distance.  The black squares represent
  the flux contained within each 1 arcmin cell of the POINT-AGAPE
  data, plotted with their equivalent disk position.  The grey points
  represent the M31 model flux evaluated for each POINT-AGAPE cell,
  again plotted with their disk position.  The contribution to the
  overall galactic light from M32 can be seen at around $a_{d}=70$
  arcmin and some structure - mainly from the dust lanes - can be seen
  within the disk across most of the outer galaxy.}
\label{light_fit}
\end{figure}

\subsection {Testing the distributions}
\label{se:b_only}

By employing the detection probability function
(Equation~\ref{probdist}) we are able to test three special cases of
the CN distribution in M31, namely that novae follow: the overall galactic light;
the bulge light only; or the disk light only.  We use the Kolmogorov-Smirnov (K-S) Test in
all three cases to ask whether the nova distribution detected is
consistent with being drawn from each of the three eruption
distributions.

\begin{figure*}
\rotatebox{270}{\includegraphics[width=230pt]{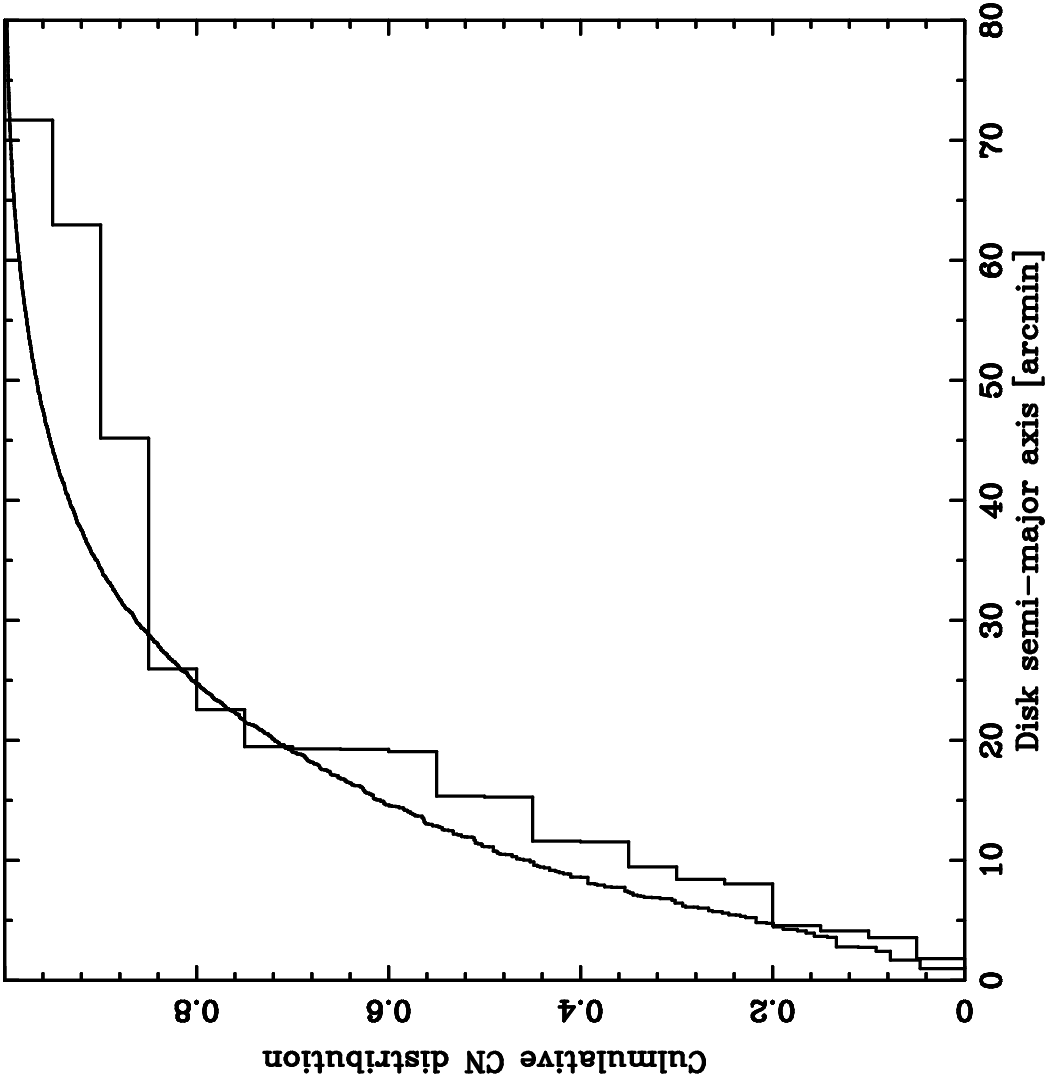}}
\rotatebox{270}{\includegraphics[width=230pt]{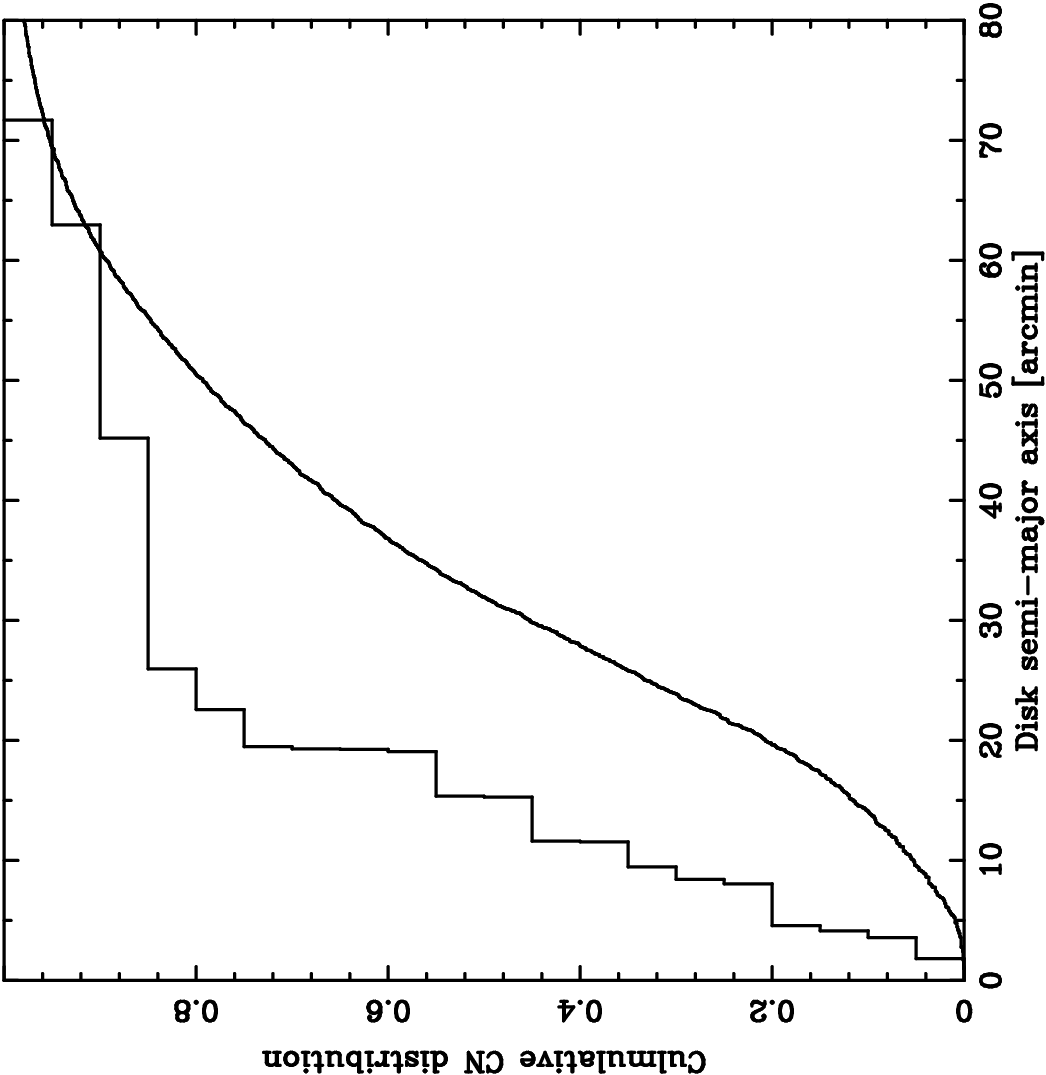}}
\rotatebox{270}{\includegraphics[width=230pt]{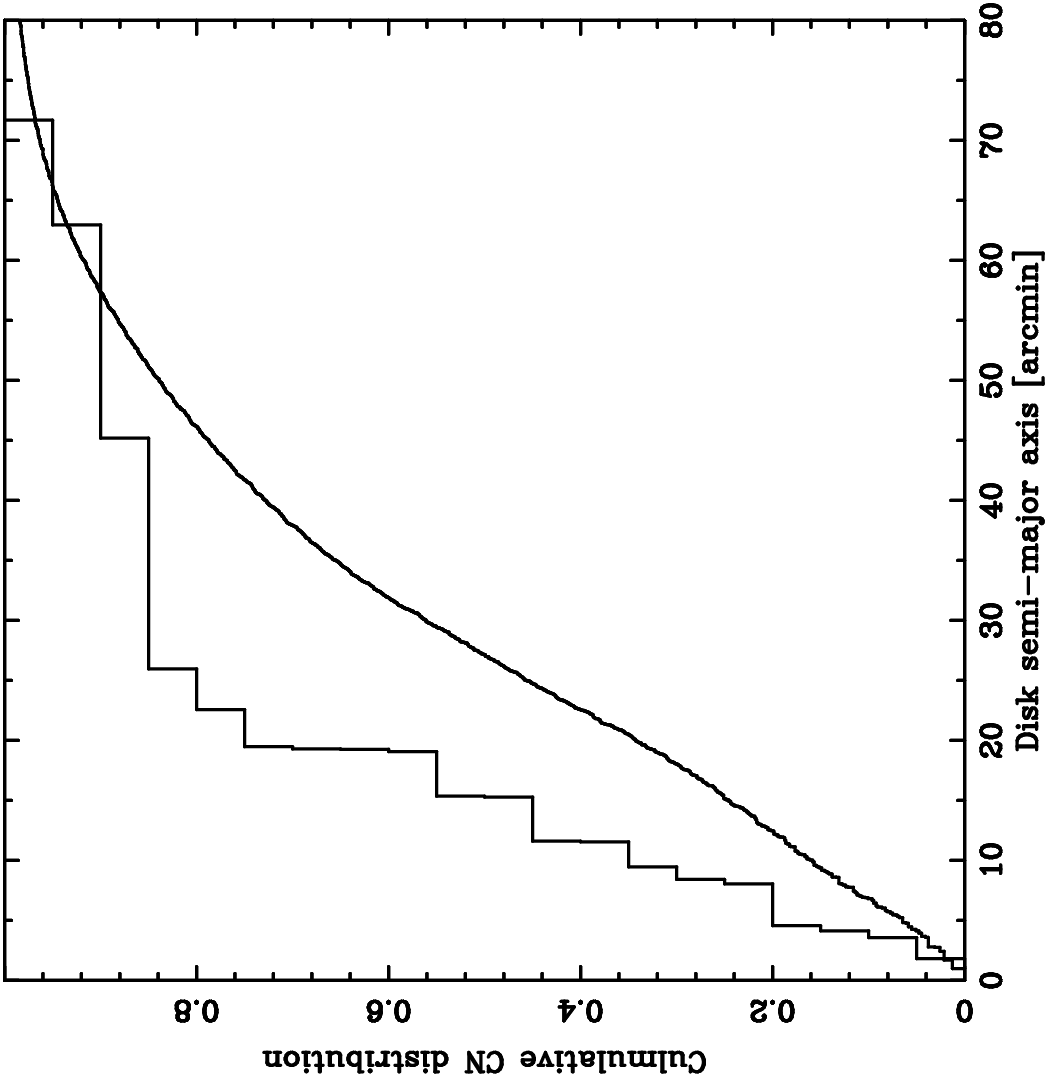}}
\rotatebox{270}{\includegraphics[width=230pt]{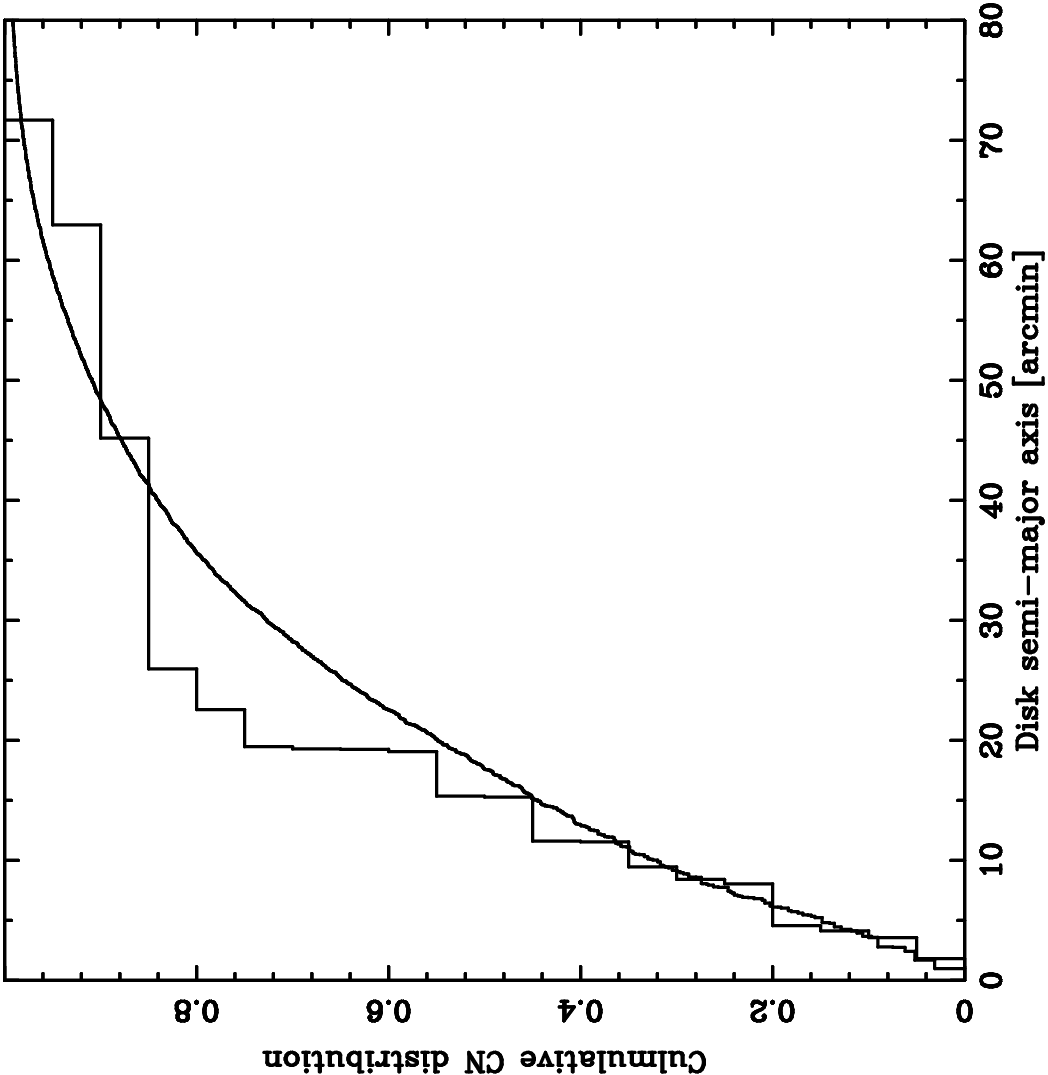}}
\caption{The major axis distribution of the 20 detected POINT-AGAPE
  novae (grey histograms) along with theoretical predictions of four models (black lines).
  The upper left panel shows the
  bulge-only model, whilst the upper right panel shows the disk-only model.
  The lower left panel shows the galactic light model, whereas the
  lower right panel represents the most probable distribution (see
  Section~\ref{se:twopop}).}
\label{ks_dist}
\end{figure*}

For the bulge-only model we use the K-S Test to determine whether the
flux model and the CN distribution are drawn from the same parent
population.  The upper left plot in Figure~\ref{ks_dist} shows the
cumulative distribution of bulge detection probability with increasing
disk semi-major axis ($a_{d}$) compared with the cumulative detected
CN distribution with increasing disk semi-major axis.  The K-S Test
produces a probability that the two distributions are drawn from the
same parent population of 0.43.  Hence, it is clear that the bulge
alone can give rise to the observed distribution of CNe.

Next we test the disk-only model. The upper right plot in
Figure~\ref{ks_dist} shows the cumulative distribution of disk
detection probability with increasing disk semi-major axis ($a_{d}$)
compared with the cumulative detected CN distribution with increasing
disk semi-major axis.  The probability that these two distributions
are drawn from the same population is $4.2\times10^{-6}$.  It is
therefore quite clear that the disk alone can not account for the
observed distribution of CNe.

Finally we test the galactic light model. The lower left plot in
Figure~\ref{ks_dist} shows the cumulative distribution of nova
detection probability with increasing disk semi-major axis $(a_{d})$,
again compared with the cumulative detected novae distribution with
increasing disk position.  There is a probability of $1.2\times10^{-3}$
that these distributions are drawn from the same population.

From the results of these tests it is clear that the bulge-only model
does a good job of reproducing the observed CN distribution, whilst
the disk-only and galactic light models do poor jobs.  However, given
that the bulge-model over-estimates the nova distribution near the
centre of M31, where the bulge dominates (see Figure~\ref{ks_dist}
upper left plot), it is clear that a combination of both bulge and
disk populations, with different weightings, is required to adequately
model the detected CN distribution.

\subsection {The two population model}
\label{se:twopop}

Following the results of the testing of the three special case CN
eruption models, it seems clear that the favoured model
comprises a combination of both disk and bulge
populations, with each population having a different
eruption rate per unit $r'$ flux.  To test this new model, we first
make the assumption that the nova eruption probability in the disk or
the bulge is proportional to the disk or bulge luminosity
respectively:
\begin{equation}
p_{i}\propto \sigma_{d}\cdot f_{i}^{d}+\sigma_{b}\cdot f_{i}^{b}
\label{final_prob}
\end{equation}
where the disk flux, $f^{d}$, and the bulge flux, $f^{b}$,
are defined in Equations~\ref{diskmodel} and \ref{bulgemodel}
respectively and $\sigma_{d}$ and $\sigma_{b}$ are the number of CN
eruptions per unit time per unit $r'$ flux for the disk and bulge
populations respectively.

In order to test this new hypothesis we define the probability of
detecting a CN in a particular cell, given that a CN both erupts and
is detected within one of the two POINT-AGAPE fields:
\begin{equation}
P_{i}=\frac{(\theta f_{i}^{d}+f_{i}^{b})\cdot\varepsilon_{i}}{\theta\sum_{j=1}^{N_{bins}}f_{j}^{d}\cdot\varepsilon_{j}+\sum_{j=1}^{N_{bins}}f_{j}^{b}\cdot\varepsilon_{j}}
\label{detect_prob_final}
\end{equation}
where $\theta$ is the ratio of disk and bulge population
eruption rates per unit $r'$ flux.

In order to constrain the favoured value of $\theta$ we employ a
maximum likelihood test.  The likelihood function chosen is shown in
Equation~\ref{mlh1} below.  This function is derived from a simple
Poisson analysis of our nova detection model, for a given underlying
mean number of expected detections, evaluated over all possible
underlying means.
\begin{equation}
P_{model}=\int_{0}^{\infty}\frac{\mu^{N}}{N!}e^{-\mu}\prod_{i=1}^{N_{bins}}\lambda_{i}^{n_{i}}e^{-\lambda_{i}}d\mu
\label{mlh1}
\end{equation}
where $\mu$ is the underlying mean number of expected
detections, $N$ is the total number of CN detected by the POINT-AGAPE
survey (20), $n_{i}$ is the number of CN detected in each data bin and
$\lambda_{i}$ is the expected number of CN detected in each bin, given
by:

\begin{equation}
\lambda_{i}=\mu\cdot\frac{(\theta
f_{i}^{d}+f_{i}^{b})\cdot\varepsilon_{i}}{\theta\sum_{j=1}^{N_{bins}}f_{j}^{d}\cdot\varepsilon_{j}+\sum_{j=1}^{N_{bins}}f_{j}^{b}\cdot\varepsilon_{j}}
\label{mlh2}
\end{equation}

In order to confine the range over which the likelihood function is
investigated we change variables from $\theta$ to the bulge fraction
($\Phi$), where $\Phi$ is defined as the fraction of the eruption
probability within the POINT-AGAPE field due to the M31 bulge:
\begin{equation}
\theta=\frac{1-\Phi}{\Phi}\cdot\frac{\sum_{j=1}^{N_{bins}}f_{j}^{b}}{\sum_{j=1}^{N_{bins}}f_{j}^{d}}
\label{mlh3}
\end{equation}

By evaluating the distribution of the likelihood function we 
derive the most likely value of $\Phi=0.67$ and, by assuming a linear prior
in $\Phi$, we evaluate confidence limits about the most likely
value.  As such, we find that the 95\% confidence interval of $\Phi$
is
\begin{equation}
0.46\leq\Phi\leq0.82
\label{mlh4}
\end{equation}
Using Equation~\ref{mlh3}, this equates to a favoured value
of $\theta=0.18$ with the 95\% confidence interval bounded by
$\theta=0.91$ and $\theta=0.02$.  This result also allows us to rule
out models with $\sigma_{d}\geq\sigma_{b}$ at the 95\% level, lending
strong support to the existence of separate bulge and disk CN
populations.

\section{M31 CN rate}\label{rate}

The CN eruption probability model is used, in conjunction with the
detection completeness data, to compute an estimate of the global nova
rate in M31.  The total number of CNe observed within the POINT-AGAPE
fields must be proportional to the total probability of detecting a CN
within those fields, so we have
\begin{equation}
\xi\sum_{i=1}^{N_{bins}}\varepsilon_{i}\cdot\Psi_{i}=n
\label{rateequation}
\end{equation}
where $\varepsilon_{i}$ is the probability of detecting an
erupted CN at a particular location, $\Psi_{i}$ is the probability of
a CN erupting at that location, $n$ is the number of novae detected
within the POINT-AGAPE dataset (20), and $\xi$ is an unknown constant
relating the detection probability to the recovered number of novae.
The definition of the eruption probability given in
Equation~\ref{final_prob} is used to define $\Psi_{i}$:
\begin{equation}
\Psi_{i}=\frac{\theta
f_{i}^{d}+f_{i}^{b}}{\theta\sum_{j=1}^{N_{bins}}f_{j}^{d}+\sum_{j=1}^{N_{bins}}f_{j}^{b}}
\label{psi_eq}
\end{equation}
The value of the multiplier $\xi$ can be computed for a
number of different values of $\theta$, thus producing a range of M31
nova rates.  However, we will restrict the values of $\theta$
examined to those that relate to specific physical situations:
$\theta=0$, the bulge-only system; $\theta=1$, the galactic light
scenario; $\theta\rightarrow\infty$, the disk-only system and
$\theta=0.18$, the favoured value produced by the maximum likelihood
analysis of the two population model.  The computed values of $\xi$
for these models are given in Table~\ref{xi_table}.

\begin{table*}
\caption{The computed values of the nova rate normalisation ($\xi$), the approximate bulge-to-disk probability
  ratio, the fraction of novae erupting within the POINT-AGAPE fields ($\varphi$), the underlying number of nova eruption ($N$) during the survey lifetime, the M31 bulge nova rate
  ($\dot{N}_{bulge}$), the disk nova rate ($\dot{N}_{disk}$) and the
  global nova rate ($\dot{N}$) for a range of different CN eruption
  probability models.}
\label{xi_table}
\begin{tabular}{llllllll}
\hline
$\theta$            & $\xi$ & Bulge:disk       & $\varphi$ & $N$  & $\dot{N}_{bulge}$ & $\dot{N}_{disk}$ & $\dot{N}$    \\
                    &       & probability      &           &      & (year$^{-1}$)     & (year$^{-1}$)    & (year$^{-1}$)\\
                    &       & ratio            &           &      &                   &                  &              \\
\hline
0.00                & 92.61 & 1:0              & 0.58      & 159  & $56\pm13$         & -                & $56\pm13$    \\
0.18                & 86.01 & 4:3              & 0.47      & 184  & $38\pm8$          & $27\pm6$         & $65\pm15$    \\
1.00                & 77.99 & 1:4              & 0.37      & 213  & $15\pm3$          & $61\pm14$        & $75\pm17$    \\
$\rightarrow\infty$ & 72.73 & 0:1              & 0.31      & 233  & -                 & $82\pm18$        & $82\pm18$    \\
\hline
\end{tabular}
\end{table*}

The global M31 CN number can now be computed from
\begin{equation}
N=\frac{\xi}{\varphi}
\label{N_eq}
\end{equation}
where $N$ is the global nova number and $\varphi$ is a
constant multiplier that accounts for the proportion of the total
galactic eruption probability that has been sampled by the POINT-AGAPE
survey.  $\varphi$ is defined by
\begin{equation}
\varphi=\frac{\sum_{i=1}^{N_{bins}}(\theta
f_{i}^{d}+f_{i}^{b})}{\sum_{i}^{M31}(\theta f_{i}^{d}+f_{i}^{b})}.
\label{phi_eq}
\end{equation}
In order to evaluate the sum over the entire galaxy (the denominator
of Equation~\ref{phi_eq}) we extend the 1 arcmin grid, initially used
to model the completeness, over all space.  As galactic disks are
known to be truncated radially at a distance of 3--4 scale lengths
\citep{1981A&A....95..105V,2000A&A...357L...1P}, we evaluated the sum
out to a distance of 20 kpc \citep{ibata2005} (equivalent to a disk
semi-major axis distance of $\sim90$ arcmin) from the centre of M31.
Table~\ref{xi_table} shows the computed values of $\varphi$ for the
models tested.  We are now able to compute the global number of CNe
that erupted in M31 during the POINT-AGAPE observing baseline.  These
values are also shown in Table~\ref{xi_table} for the four model
examples.

However, to calculate the global M31 CN rate, we first need to take
account of the finite observable lifespan (as defined solely by our
observations) of each of the detected novae.  As we did not require
that a nova's light-curve should be completely contained within our
data, our effective baseline for each CN is extended by the lifespan
of that particular CN.  As the novae have been seeded uniformly over
the POINT-AGAPE fields and the seeded nova is selected randomly, the
baseline of observations has been extended, on average, by the mean
lifetime of all 20 novae:
\begin{equation}
T=T_{baseline}+\bar{t}_{nova}
\label{time_extend}
\end{equation}
where, $T_{baseline}$ is the time between the first and last
POINT-AGAPE observation (2.472 years), and $\bar{t}_{nova}$ is the mean
lifetime of the 20 POINT-AGAPE novae (0.359 years).  The effective
baseline for CNe of the POINT-AGAPE survey is therefore 2.830 years.
The nova rate for M31 is then
\begin{equation}
\dot{N}=\frac{\xi}{\varphi T}.
\label{nova_rate}
\end{equation}
The computed M31 global CN rates for our four model
scenarios are given in Table~\ref{xi_table}.  This illustrates that
the predicted overall nova rate is only modestly dependent upon the
eruption model chosen.  However, the separate bulge and disk novae
rates show a strong dependence upon $\theta$ as expected.

The errors shown for the nova rates in Table~\ref{xi_table} are
generated solely from the Poisson errors related to the size of our
nova catalogue.  Given the small size of this catalogue, this source
of error ($\sim22\%$) is expected to dominate over all others.  The
other main sources of error arise from the completeness
calculations, the lack of fast novae in the catalogue, the possible
misidentification of novae and the modelling of the surface
brightness.  In addition, our limited knowledge of the internal
extinction of M31 makes it difficult to estimate the errors introduced
into the completeness by its exclusion.  However, we expect these
errors to be small.  The maximum $r'$ extinction expected in the disk
is $\sim0.7$ magnitudes; as all the POINT-AGAPE novae were followed
through at least one magnitude, it is expected that few, if any, novae
were missed due to extinction problems.  Should a non-CN have been
wrongly included within the nova catalogue, this would directly result
in a 5\% reduction of the nova rate (see Equation~\ref{rateequation}),
coupled, indirectly, with a further maximum decrease of 5\% from the
completeness model.  Hence, the misidentification of a single nova
generates a maximum absolute error of $\sim7\%$.  However, the errors
introduced into the completeness through misidentification are highly
likely to be much less than 5\%.  Likewise, a single CN which is
``missed'' due to extinction would induce a maximum increase of
$\sim7\%$ of the global nova rate.  Although there is some error in
our modelling of the M31 surface brightness within the POINT-AGAPE
fields, the good fit of the model shown in Figure~\ref{light_fit}
suggests that this is modest.

Taking the above error discussion into account, we can use the value
of $\theta=0.18^{+0.24}_{-0.10}$ deduced from the likelihood analysis
of the two population model to produce a model constrained estimate of
the true nova rate of M31.  By evaluating the error on the bulge
fraction determination we are able to show that the Poisson errors are
still the dominating error source although there is also a
significant contribution from the uncertainty in the model.  Hence we
arrive at the following estimate of the true observable nova rate of
M31:
\begin{equation}
\dot{N}=65^{+16}_{-15}\;\mathrm{year}^{-1}
\label{truerate}
\end{equation}

\section{Discussion and conclusions}\label{conclusion}

\subsection {Novae as distance indicators}

Sections~\ref{mmrd} and \ref{15} report the calibration of both the
MMRD and $t_{15}$ relationships using the POINT-AGAPE CN catalogue,
which includes a recalibration of each nova's decay rate and speed
class and assessments of the uncertainties in the maximum light and
line of sight extinction to each nova.

The two POINT-AGAPE MMRD relationships (see Equations~\ref{calcMMRD3b}
and \ref{calcMMRD3b_i}) are consistent with the existence of an MMRD
relationship for the $r'$ and $i'$ filters.  In fact, the observed
scatter in both relationships, although higher than that of recent
Galactic calibrations, can be accounted for solely by extinction and
maximum light uncertainties.  We are able to show that, for the speed
classes used for the MMRD calibration, missing the maximum light of a
nova by up to a week is not the dominating factor in the MMRD scatter.
We also find that for the bright novae for which a linear MMRD is
expected, extinction corrections make little difference to the slope
determination. However, it is clear that a better understanding of the
extinction affecting the POINT-AGAPE novae is required in order to
make a more precise calibration of the MMRD within M31.  Little more
can be said about the comparison between previous MMRD relationship
calibrations and the POINT-AGAPE calibrations, as the Galactic MMRD
(and previous M31 relations) are calibrated using bluer filter bands
than the POINT-AGAPE filters.  In fact our calibrations constitute the
first attempt to do so using Sloan filters.  Given that CNe become
bluer as they decline, we would naively expect the POINT-AGAPE $r'$
and $i'$ slopes to be steeper than the galactic $V$-band slope,
whereas we find that the $r'$ slope is remarkably similar, with the
$i'$ slope being much steeper, as expected.  However, it should also
be noted that the $r'$ filter contains the H$\alpha$ emission line.
As CNe are known to remain bright in H$\alpha$ long after the visible
light-curve has diminished, this may be adversely increasing our
measured $r'$ decline times.  It is also known that the decline of the
H$\alpha$ emission of a CN is not well correlated with its H$\alpha$
luminosity at maximum \citep{2005clno.book...14S}. As such, this could
potentially detract from the usefulness of any $r'$ MMRD relationship
(Shafter, private communication).

The MMRD relationship may be be used as a tool to measure the relative
distance between two populations of novae.  However, given that our
calibrations are the first to be carried out for the Sloan $r'$ and
$i'$ bands, it would be inappropriate to attempt to estimate the M31
distance by comparison with Galactic $V$ and $B$ band relationships.

The analysis of the $r'$ and $i'$ $t_{15}$ relationships within the
POINT-AGAPE catalogue is, like the MMRD relationship, dominated by the
extinction uncertainties within the data.  However, the extent of the
scatter observed in both the $r'$ and $i'$ data can not be accounted
for by the extinction and maximum light uncertainties alone.  A
comparison of our $t_{15}$ values with those for bluer bands are,
however, consistent with a CN becoming bluer following maximum light.

Figure~\ref{t15_scatter} shows the distribution of $t_{n}$ scatter,
$n$ days after maximum light.  This plot clearly indicates that the
scatter between the light-curves is large over the entire period
sampled.  Also it is clear that there is no evidence of a minimum in
the scatter for times around 15 days.  The small minima at $\sim30$
days in the $r'$ data and $\sim35$ days for the $i'$ are related to
the sampling of the surveys.

To some extent, the $t_{15}$ analysis is limited by the temporal
sampling of the POINT-AGAPE survey.  Unlike the MMRD relationship that
requires good sampling around the peak of the light-curve and some
good sampling of the subsequent decline, to test the $t_{15}$
relationship one also requires good sampling of the light-curve
specifically at $\sim15$ days after peak.  Due to the make up of the
survey, the light-curves are generally constructed from short periods
of good sampling, followed by regions with no data (see Figure~2 of
Paper~1).  As such, a relatively large amount of extrapolation is
required to estimate each nova's flux between observations.  Given the
rather erratic behaviour of a CN's light-curve, the estimation of the
errors induced by linearly interpolating over large periods with no
data is a far from trivial task.  The $r'$ data are again likely to be
adversely affected by the H$\alpha$ emission.  We can conclude that
the POINT-AGAPE CN catalogue shows no evidence of a $t_{15}$
relationship, nor strong evidence of convergence at another timescale.
We would require nova light-curves with much more uniform sampling
than the POINT-AGAPE novae to be able to make a more definite
statement regarding the $t_{n}$ relationship's overall validity in the
$r'$ and $i'$ filters and its potential usefulness.

\subsection{Completeness}

Overall, the method employed to evaluate the completeness of the
POINT-AGAPE CN catalogue generated by the nova pipeline allowed us to
obtain a very good understanding of the CN detection efficiency of
both the survey and the pipeline.  The completeness analysis took into
account a variety of possible selection effects which prevent us from
detecting novae.  These selection effects included the strongly
varying surface brightness of M31, the range of morphologies exhibited
by CN light-curves and the temporal sampling of the POINT-AGAPE
survey.

Until very recently the detection of novae relied solely upon visual
detection, often by the ``blinking'' of images.  Even the most recent
surveys \citep[for example]{2001ApJ...563..749S} have relied on some
visual inspection, particularly to aid in the detection of the
faintest novae.  The majority of past nova surveys have also relied
upon visual inspection of light-curves to determine the likelihood
that an object was a CN \citep[e.g][]{2003ApJ...599.1302F}.  As our
pipeline uses much more robust methods and objective selection
criteria to both detect and classify potential CNe, we are confident
that the completeness of the catalogue is well understood. Whilst the
POINT-AGAPE CN catalogue may not be complete\footnote{A number of CN
candidates contained within the POINT-AGAPE dataset are known not to
be contained within our catalogue
\citep{2004ApJ...601..845A,feeney2005} for reasons we understand
\citep{feeney2005}.}, we are nonetheless able to quantify our
completeness.

There were, however, a number of factors that have not been taken into
account by the completeness analysis.  As was discussed in
Section~\ref{extinction}, our knowledge of the internal extinction of
M31 is limited and these extinction uncertainties have not been built
into the completeness computations.  Whilst the extinction may be
diminishing our ability to detect novae, especially fainter CNe, its
relatively small magnitude should not be too troublesome.  There are
no very fast novae ($t_{2}\leq10$ days) within the POINT-AGAPE
catalogue\footnote{PACN-99-07 has a great uncertainty in its speed
class assignment.  Although initially classified as a very fast nova,
it is thought more likely to be a moderately fast or slow nova.}.
Whilst this may simply be indicative of our small sample size, there
may also be additional selection effects -- due to the temporal
sampling of the POINT-AGAPE survey -- that are preventing us from
detecting novae of this class.  Both of these effects potentially
prevent us from observing novae erupting during the survey.
Consequently, the computed completeness is likely to be an
over-estimate.  Given the form of the extinction within M31, it is
also likely that the completeness has been over estimated to a greater
extent within the disk (due to its generally greater extinction) than
within the bulge.  However, \citet{2001ApJ...563..749S} used
observations of the planetary nebula population of M31 to conclude
that (in H$\alpha$) the observed CN population is not significantly
affected by extinction.  As it is expected that fast novae are more
likely to be observed within the disk of a galaxy (see
Section~\ref{populations}), the probable exclusion of very fast novae
from the catalogue  again leads us to conclude that, if anything, we
have over-estimated the completeness and  so under-estimated the nova
rate.

\subsection {M31's CNe population}

The analysis of the observed CN distribution within M31 allows us to
develop a basic model of the underlying CN distribution.  Following
the separation of the disk and bulge through modelling of the
surface brightness, we are able to show that, within the two
POINT-AGAPE fields, the observed CN distribution does not follow that
of the galactic light. Nor is the observed distribution likely to
arise solely from the disk.  Although the bulge alone can support the
observed distribution, a combination of a bulge and disk population
seems required to fully reproduce the observed distribution.

\begin{figure}
\rotatebox{270}{\includegraphics[height=230pt]{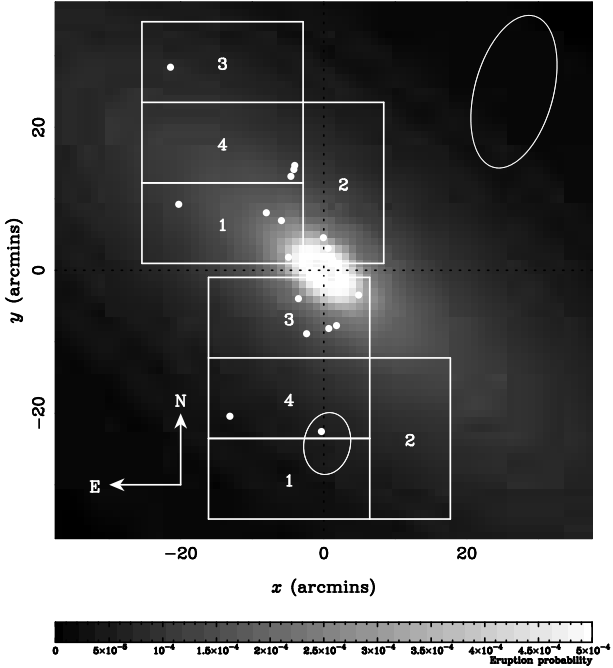}}
\caption{The favoured M31 eruption probability
  model ($\theta=0.18$), over plotted with the position of the
  POINT-AGAPE fields and the 20 detected CNe.}
\label{final_model}
\end{figure}

A maximum likelihood analysis of the two population model (see
Section~\ref{se:twopop}) indicates that the ratio of the disk and bulge
population eruption rates per unit $r'$ flux is 0.18.  This result is
consistent with previous findings
\citep{1987ApJ...318..520C,1989AJ.....97.1622C,2001ApJ...563..749S}
which reported that the M31 novae are primarily associated with the
bulge.  \citeauthor{2001ApJ...563..749S} also reported an eruption
rate per unit $B$ flux within the bulge of up to an order of magnitude
greater than that of the disk.  Figure~\ref{final_model} shows a
schematic plot of our ``best'' CN eruption model.

The range of bulge-to-disk eruption rate ratios that are consistent
with the range of observed CN distributions
($\theta=0.18^{+0.24}_{-0.10}$) leads to a range of expected
bulge-to-disk nova eruption rate ratios.  The global bulge-to-disk CN
ratios range from $5:1$ for a bulge dominant population ($\theta=0.1$)
through to $1:4$ for the distribution following the surface brightness
($\theta=1$).  The expected nova ratios within the POINT-AGAPE fields
themselves range from $20:1$ ($\theta=0.1$) to $2:1$ ($\theta=1$).
However, the POINT-AGAPE survey of M31 covers a much greater surface
area of M31 than all previous nova surveys \citep[see][for a
summary]{2005clno.book...14S}, which have concentrated mainly on the
bulge.  The POINT-AGAPE survey has given us much better coverage of
the M31 disk and its CN population.  As a result, these previous
surveys will have all observed a distribution that appears to be much
more bulge dominated than that of the POINT-AGAPE catalogue.

The analysis of the M31 CN distribution is, however, limited by a
number of considerations.  These are mainly the small size of the
POINT-AGAPE CN catalogue, the simplistic nature of the M31 surface
brightness models and the uncertainties arising from the completeness
modelling.  Further, the CN distribution analysis relies upon the
modelling of the M31 surface brightness.  However, this modelling only
includes the ``normal'' disk component and the bulge component of the
surface brightness.  Other components, such as the spiral arm
structure, the dust lanes (and extinction within M31 in general) and
M32, are not taken into account.  As with the completeness analysis,
these effects are expected to have greater affect within the disk than
the bulge.  Thus, the inclusion of the dust lanes and spiral structure
within the models could lead to an increase in the expected number of
disk novae.

\subsection {The M31 and Milky Way nova rates}

By extending the M31 CN eruption model over the whole galaxy we are
able to produce an estimate of the global nova rate.  The computed
global observable nova rate of M31 is $65^{+16}_{-15}$ year$^{-1}$,
with a bulge rate of $38^{+15}_{-12}$ year$^{-1}$ and a disk rate of
$27^{+19}_{-15}$ year$^{-1}$.  This result is at the limit of being
consistent with that of the most robust previous calibration, which
found a global rate of $37_{-8}^{+12}$ year$^{-1}$
\citep{2001ApJ...563..749S}.  However, our results are much higher
than all previous results, including the
\citeauthor{2001ApJ...563..749S} determination and those of
\citet{1929ApJ....69..103H} ($\sim30$ year$^{-1}$),
\citet{1956AJ.....61...15A} ($24\pm4$ year$^{-1}$) and
\citet{1989AJ.....97.1622C} ($29\pm4$ year$^{-1}$).  The ratio between
the bulge and disk nova rate is also markedly lower than that computed
by \citeauthor{2001ApJ...563..749S}'s (from their maximum likelihood
analysis of the M31 CN distribution).

Despite its apparent high value, we are confident that our computed
nova rate is a true evaluation of the nova production rate of M31.
Our robust completeness analysis and objective selection criteria lead
us to believe that the completeness of previous surveys may have been
over estimated.  Also, given their bulge-centric nature, many previous
surveys are likely to have under estimated the contribution from disk
novae.  Sources of concern in our estimated rate arise again from the
extinction uncertainties and the lack of very fast novae in the
catalogue.  However, both these factors potentially lead to a further
increase in the predicted rate, so we are led to conclude that the
true global CNe rate of M31 is higher than was previously thought by
around $50\%$.

We can use our estimated global M31 nova rate, along with our computed
eruption rates for the bulge and the disk, to produce an estimate of
the global Galactic nova rate.  Using a similar method to that
outlined in \citet{2002cne..conf..462S}, we assume a Galactic
disk-to-bulge luminosity ratio of $\sim8$ and a Milky Way to M31
luminosity ratio of 2/3.  Hence, we compute a global Galactic nova
rate of $34^{+15}_{-12}$ year$^{-1}$, with a disk rate of
$20^{+14}_{-11}$ year$^{-1}$ and a bulge rate of $14^{+6}_{-5}$
year$^{-1}$.  These rates are broadly consistent with other Galactic
estimates based on M31 data and broadly consistent with estimates
based upon direct observations of Galactic novae
\citep{2002cne..conf..462S}.

\section*{Acknowledgements}

MJD and MW are supported by studentships from the Particle Physics and
Astronomy Research Council (PPARC). EJK and MFB are supported,
respectively, by a PPARC Advanced Fellowship and a PPARC Senior
Fellowship. SCN is supported by the Swiss National Science
Foundation. JA and YT are supported by the Leverhulme Trust.

The authors are grateful to Chris Collins for useful discussions on
the statistical analysis.

\bibliographystyle{mn2e} \bibliography{refs}

\end{document}